\documentclass[conference]{IEEEtran}
\IEEEoverridecommandlockouts
\usepackage{cite}
\usepackage{amsmath,amssymb,amsfonts}
\usepackage{graphicx}
\usepackage{textcomp}
\usepackage{xcolor}
\usepackage{empheq}
\usepackage{algorithmicx}
\usepackage{algorithm,algpseudocode}
\usepackage{algorithmicx}
\usepackage{setspace}
\usepackage{subcaption}
\usepackage{siunitx}
\usepackage{amsmath}
\usepackage{float}
\usepackage{color}
\usepackage{url}
\usepackage{dblfloatfix} 
\usepackage{graphicx}
\usepackage{verbatim}

\usepackage{multirow}
\usepackage[short]{optidef}
\usepackage{balance}
\usepackage{hyperref}
\hypersetup{
    colorlinks=true,
    linkcolor=blue,
    filecolor=magenta,      
    urlcolor=cyan,
}

\def\BibTeX{{\rm B\kern-.05em{\sc i\kern-.025em b}\kern-.08em
    T\kern-.1667em\lower.7ex\hbox{E}\kern-.125emX}}
\begin{document}

\title{RF SSSL by an Autonomous UAV with Two-Ray Channel Model and Dipole Antenna Patterns
\thanks{This research is supported in part by the NSF award CNS-1939334.}
}

\author{\IEEEauthorblockN{Hyeokjun Kwon, Sung Joon Maeng, and Ismail Guvenc
}
\IEEEauthorblockA{Department of Electrical and Computer Engineering,
North Carolina State University, Raleigh, NC}
\IEEEauthorblockA{\{khyeokj,smaeng,iguvenc\}@ncsu.edu}
}

\maketitle

\begin{abstract}
Advancements in unmanned aerial vehicle (UAV) technology have led to their increased utilization in various commercial and military applications. One such application is signal source search and localization (SSSL) using UAVs, which offers significant benefits over traditional ground-based methods due to improved RF signal reception at higher altitudes and inherent autonomous 3D navigation capabilities. Nevertheless, practical considerations such as propagation models and antenna patterns are frequently neglected in simulation-based studies in the literature. In this work, we address these limitations by using a two-ray channel model and a dipole antenna pattern to develop a simulator that more closely represents real-world radio signal strength (RSS) observations at a UAV. We then examine and compare the performance of previously proposed linear least square (LLS) based localization techniques using UAVs for SSSL. Localization of radio frequency (RF) signal sources is assessed based on two main criteria: 1) achieving the highest possible accuracy and 2) localizing the target as quickly as possible with reasonable accuracy. Various mission types, such as those requiring precise localization like identifying hostile troops, and those demanding rapid localization like search and rescue operations during disasters, have been previously investigated. In this paper, the efficacy of the proposed localization approaches is examined based on these two main localization requirements through computer simulations.
\end{abstract}


\section{Introduction}
The availability of unmanned aerial vehicles (UAVs) has expanded in various industries, leading to rapid growth in the UAV market. As interest in the UAV industry continues to increase, numerous studies and experiments are being conducted aggressively. Especially compared to ground systems, UAVs have significant advantages in achieving line of sight. In addition, UAVs offer flexibility in deployment for various situations with immediate response time. These characteristics make UAVs well-suited for signal source searching and localization (SSSL) missions. 

Requirements for SSSL can vary in different situations. From a high-level perspective, we can categorize the UAV localization requirements into two: 1) localization accuracy with fixed flight time (LAFFT): requiring a high localization accuracy, and 2) localization time with fixed localization accuracy (LTFLA): demanding fast localization with a coarse accuracy. For example, the implementation of UAVs for positioning hostile targets with RF signals in warfare is studied in \cite{military,hanguk}. This is a representative example of LAFFT because accurate localization is highly critical for a precise strike in warfare. On the other hand, the localization, search, and rescue of victims (e.g., after disasters) in an undetermined vast area within a limited time is studied in \cite{Oh2021, Atif2021}. This can be categorized as LTFLA since the localization time may be more critical than achieving a very low localization accuracy. 

Estimating an RF signal location using UAVs is covered in various studies in the existing literature~\cite{Güzey2022,bhattacherjee2022experimental,Hasanzade2018,Annepu2020}. 
%Various factors are defined as determinants for the accurate or fast estimation of RF signals. For example, 
The importance of optimizing UAV search patterns for improving localization accuracy is explored in \cite{Demiane2020}, while in \cite{Liu2019, Ma2018}, the effect of the unknown path loss exponent on localization performance is studied. The dependence of the localization accuracy on the transmitter and receiver's antenna patterns are quantified in~\cite{sinha2022impact}. 
%One remarkable factor in the previous studies is the propagation channel model. 
Most previous studies use the free space path loss model in an outdoor environment as a propagation model because of its simplicity. 

In this paper, we utilize UAV-based SSSL algorithms suggested in previous work \cite{kwon2023rf}, using more realistic propagation models and antenna patterns. 
%However, as we update the system models by adjusting the environmental factors to be closer to reality, a more reality-oriented simulation setup becomes available. In this context, 
The major contributions of this paper can be summarized as follows. First, rather than the free-space propagation assumption in~\cite{kwon2023rf}, we use the two-ray propagation model that may to some extent characterize air-to-ground propagation in an open field. Moreover, we consider both omnidirectional and doughnut-shaped dipole antenna patterns with different UAV altitudes for performance analysis. We evaluate a modified version of the SSSL algorithms in~\cite{kwon2023rf} and compare their localization accuracy and localization time for various scenarios. 

The remainder of this paper is organized as follows. In Section \ref{Sec:SysModel}, two-ray propagation models, antenna gain models, and the least square-based localization approach are discussed. In Section~\ref{Sec:PropScheme}, three different SSSL approaches  are reviewed along with  the new simulation setup. In Section \ref{Sec:NumResult}, experimental configurations and setups are presented, and localization performance upon antenna patterns and UAV's altitude are compared. Consequently, the conclusion and future works are suggested in Section \ref{Sec:Conclusion}.

\section{System Model}\label{Sec:SysModel}
The coordinates of a transmitter that is to be localized are represented as $l^T = (x, y, z)$. The receiver's (UAV's) location at the $i_{\rm th}$ discretized coordinate is expressed as $l^D = (x_{i}, y_{i}, z_{i})$. We assume that the UAV altitude is fixed for a given SSSL mission, hence, $z_i=h$. 
%In the present study, a comprehensive system model has been developed, taking into account the two-ray propagation as well as antenna patterns. 
Considering a two-ray propagation model, the received signal strength can be characterized as follows \cite{maeng2018aeriq}:
\begin{equation}\label{eq:1}
{\rm P}_{\rm r} = {\rm P}_{\rm t}\left( \frac{\lambda}{4\pi}\right)^{2} \left| \frac{\sqrt{{\rm G}(\theta_{l})}}{d_{\rm los}} + \Gamma(\theta_{r})\sqrt{{\rm G}(\theta_{r})}\frac{e^{-{\rm j}\triangle\varnothing}}{d_{\rm ref}}\right|^2~,
\end{equation}
where ${\rm P}_{\rm r}$ is received signal strength, ${\rm P}_{\rm t}$ is transmitted signal strength, $\lambda$ is $\frac{f}{c}$, $\theta_{l} = \tan^{-1}(\frac{z-h}{d_{\rm 2D}})$ is the elevation angle between the transmitter and the receiver and $\theta_{r} = \tan^{-1}(\frac{z+h}{d_{\rm 2D}})$ is the ground reflection angle. To calculate $\theta_{l}$ and $\theta_{r}$, we use $d_{\rm 2D}=\sqrt{(x-x_{i})^2 + (y-y_{i})^2}$. Moreover, ${\rm G}(\theta_{l})$ is the antenna gain of the line of sight and ${\rm G}(\theta_{r})$ is the antenna gain of the ground reflected ray. In this context, the antennas of both the transmitter and the receiver are oriented vertically. Based on these assumptions, the antenna gains can be represented as a function of the spatial coordinates pertaining to the transmitter and the receiver in the following manner \cite{maeng2020interference}:
\begin{equation}\label{eq:2}
{\rm G}(\theta) = {\rm G}(l^T, l^D) = \frac{\cos(\frac{\pi \iota f}{c}\sin(\theta))-\cos(\frac{\pi \iota f}{c})}{\cos(\theta)}~,
\end{equation}
where $\iota = \frac{\lambda}{2}$ is the wavelength that is assumed to be a half-wave dipole antenna in this paper.
\begin{figure}[t!]
%\vspace{-1mm}
\centering
\includegraphics[trim=0.2cm 0.1cm 0.2cm 0.3cm, clip,width=7.8cm]{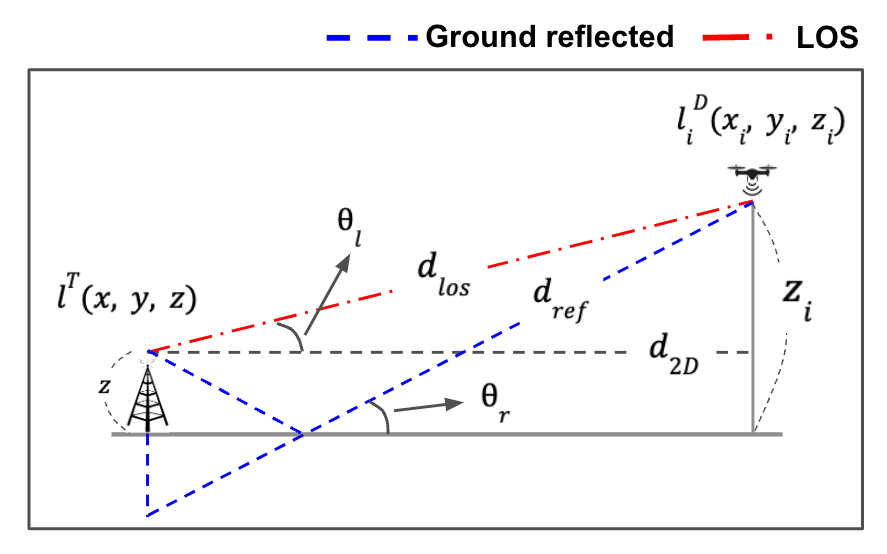}
\caption{Two-ray propagation model.}
\label{fig:tworay}
%\vspace{-5mm}
\end{figure}
In \eqref{eq:1}, $\rm {\Gamma}(\theta)$ is the ground reflection coefficient and it is represented as \cite{uavcom}
\begin{equation}\label{eq:3}
\rm {\Gamma}(\theta_{r}) = \frac{\sin(\theta_{r}) - \sqrt{\varepsilon_{0}-\cos^2(\theta_{r})}}{\sin(\theta_{r}) + \sqrt{\varepsilon_{0}-\cos^2(\theta_{r})}}~,
\end{equation}
where $\varepsilon_{0} = \varepsilon-\rm{j}60\Psi\lambda$ is an environment constant that represents the relative permittivity of the ground. This value depends on the two components of ground type: i) $\Psi$ represents the conductivity of the earth (units in mhos per meter), and ii) $\varepsilon$ denotes the dielectric constant of the ground relative to unity in free space. 

In \eqref{eq:1}, $\triangle\varnothing$ is the phase difference between the line of sight and the reflected paths by the different arrival time, which can be expressed as
\begin{equation}\label{eq:4}
\triangle\varnothing = \frac{2\pi\left( d_{\rm ref} - d_{\rm los}\right)}{\lambda}~,
\end{equation}
where $d_{\rm los}$ and $d_{\rm ref}$ are the line of sight distance between the transmitter and the $i_{\rm th}$ receiver and the reflected distance from each, respectively. They can be expressed as functions of $l^T$ and $l^D$ as follows
\begin{equation}\label{eq:5}
d_{{\rm los}_i}(l^T, l^D) = \sqrt{\left( x - x_{i} \right)^2 + \left( y - y_{i} \right)^2 + \left( z - h \right)^2}~,
\end{equation}

\begin{equation}\label{eq:6}
d_{{\rm ref}_i}(l^T, l^D) = \sqrt{\left( x - x_{i} \right)^2 + \left( y - y_{i} \right)^2 + \left( z + h \right)^2}~.
\end{equation}
We assume that the transmitted signal strength of the target is known by the UAV. Then, the path loss at each receiver's location can be modeled by  
\begin{equation}\label{eq:7}
{\rm PL}_{i} = {\rm P}_{\rm t} - {\rm P}_{{\rm r}_{i}} + \omega_{i}~,
\end{equation}
where $\omega$ denotes shadowing component that follows a zero-mean Gaussian distribution $\omega \sim N(0, \sigma^{2})$. Then, the measured path loss by the received signal strength at the $i_{\rm th}$ location can be expressed as
\begin{equation}\label{eq:8}
\widetilde{{\rm PL}_{i}} = {\rm P}_{\rm t} - {\rm P}_{{\rm r}_{i}}~.
\end{equation}
To implement the least square algorithm for localization, the distance between the target and the $i_{\rm th}$ receiver's locations needs to be estimated. We assume that the two-dimensional separation between the transmitter and the receiver can be estimated based on the residuals between the measured path loss and the path loss by the analytical model. The estimated distance based on the difference in path loss can be formulated as follows
\begin{equation}\label{eq:9}
\hat{d}_{\rm 2D} = \arg \min_{d}~\Big(\widehat{{\rm PL}_{i}} - \widetilde{{\rm PL}_{i}}\Big)^2~,
\end{equation}
where  $\hat{d}_{\rm 2D}$ represent the estimated distance between the target and the $i_{\rm th}$ UAV location in two dimensions, $\widetilde{{\rm PL}_{i}}$ denote the measured path loss, and $\widehat{{\rm PL}_{i}}$ indicates the analytically derived path loss using \eqref{eq:1}. Note that $\widehat{{\rm PL}_{i}}$ is a function of a single variable $d_{\rm 2D}$ since the parameters of this equation, ${\rm G}(\theta)$, $\Gamma(\theta{r})$, $d_{\rm los}$, and $d_{\rm ref}$, can be expressed in terms of $d_{\rm 2D}$.

Assuming that the index of the current location of the UAV is denoted as $\tilde{N}$, we only utilize the subset of indices to estimate the target location. The size of the selected unique indices to be used in the localization process should be $3\leq S\leq \tilde{N}$. The subset of indices can be represented as $\boldsymbol{\upsilon}$. In addition, a reference index $r_{\rm th}$, $r \in \boldsymbol{\upsilon}$, is defined for each suggested localization algorithm, which is used to obtain $|S-1|$ linear equations from $S$ expressions. In this paper, the closest index among the indices $\boldsymbol{\upsilon}$ from the target is set as a reference index. Once the estimated distance has been determined, the transmitter's position can be approximated by employing the least square approach for $k=1,\cdots,{S}$,~and~$k\neq r$, as~\cite{kwon2023rf}
\begin{equation}\label{eq:10}
\boldsymbol{A}_{\boldsymbol{\upsilon}}\boldsymbol{l_{\rm 2D}^{T}} = \boldsymbol{b}_{\boldsymbol{\upsilon}}~,
\end{equation}
where
\begin{align}\label{eq:11}
\boldsymbol{A}_{\boldsymbol{\upsilon}} &= 2
\begin{bmatrix}
x_{{\boldsymbol{\upsilon}(1)}}-x_{r} & & y_{{\boldsymbol{\upsilon}(1)}}-y_{r} \\
 & \vdots & \\
x_{{\boldsymbol{\upsilon}(k)}}-x_{r} & & y_{{\boldsymbol{\upsilon}(k)}}-y_{r} \\
 & \vdots & \\
x_{{\boldsymbol{\upsilon}(S)}}-x_{r} & & y_{{\boldsymbol{\upsilon}(S)}}-y_{r} 
\end{bmatrix}~,\\
%\end{align}
%and
%\begin{equation}
\label{eq:12}
\boldsymbol{b}_{\boldsymbol{\upsilon}} &= 
\begin{bmatrix}
d_{r}^2 - d_{{\boldsymbol{\upsilon}(1)}}^2 + x_{{\boldsymbol{\upsilon}(1)}}^2 + y_{{\boldsymbol{\upsilon}(1)}}^2 - \chi \\
\vdots \\
d_{r}^2 - d_{{\boldsymbol{\upsilon}(k)}}^2 + x_{{\boldsymbol{\upsilon}(k)}}^2 + y_{{\boldsymbol{\upsilon}(k)}}^2 - \chi \\
\vdots \\
d_{r}^2 - d_{{\boldsymbol{\upsilon}(S)}}^2 + x_{{\boldsymbol{\upsilon}(S)}}^2 + y_{{\boldsymbol{\upsilon}(S)}}^2 - \chi
\end{bmatrix}~,
\end{align}
and $\boldsymbol{l_{\rm 2D}^{T}}=[\hat{x},\:\hat{y}]^T$ is the estimated target location, where $\boldsymbol{\upsilon}(k)$ is $k_{\rm th}$ index of  $\boldsymbol{\upsilon}$, $d_{{\boldsymbol{\upsilon}(k)}}$ is estimated distance between the target and the receiver gained based on \eqref{eq:9}, and $\chi$ is ($x_{r}^2 + y_{r}^2$). The estimated target location $\boldsymbol{l_{\rm 2D}^{T}}$ can then be determined employing the least square solution, given by 
\begin{equation}\label{eq:13}
\boldsymbol{l_{\rm 2D}^{T}}= (\boldsymbol{A}_{\boldsymbol{\upsilon}}^T\boldsymbol{A}_{\boldsymbol{\upsilon}})^{-1}\boldsymbol{A}_{\boldsymbol{\upsilon}}^T\boldsymbol{b}_{\boldsymbol{\upsilon}}~.
\end{equation}

%Proposed Schemes
\section{Proposed SSSL Methods}\label{Sec:PropScheme}
\begin{figure}[!t]
\centering
\begin{subfigure}{0.74\columnwidth}
\centering
\includegraphics[width=\textwidth]{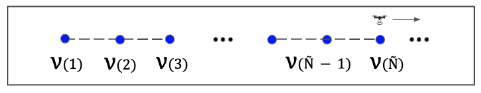}
\caption{LLS-CUM}
\label{fig:cum}
%\vspace{-1mm}
\end{subfigure}
\begin{subfigure}{0.74\columnwidth}
\centering
\includegraphics[width=\textwidth]{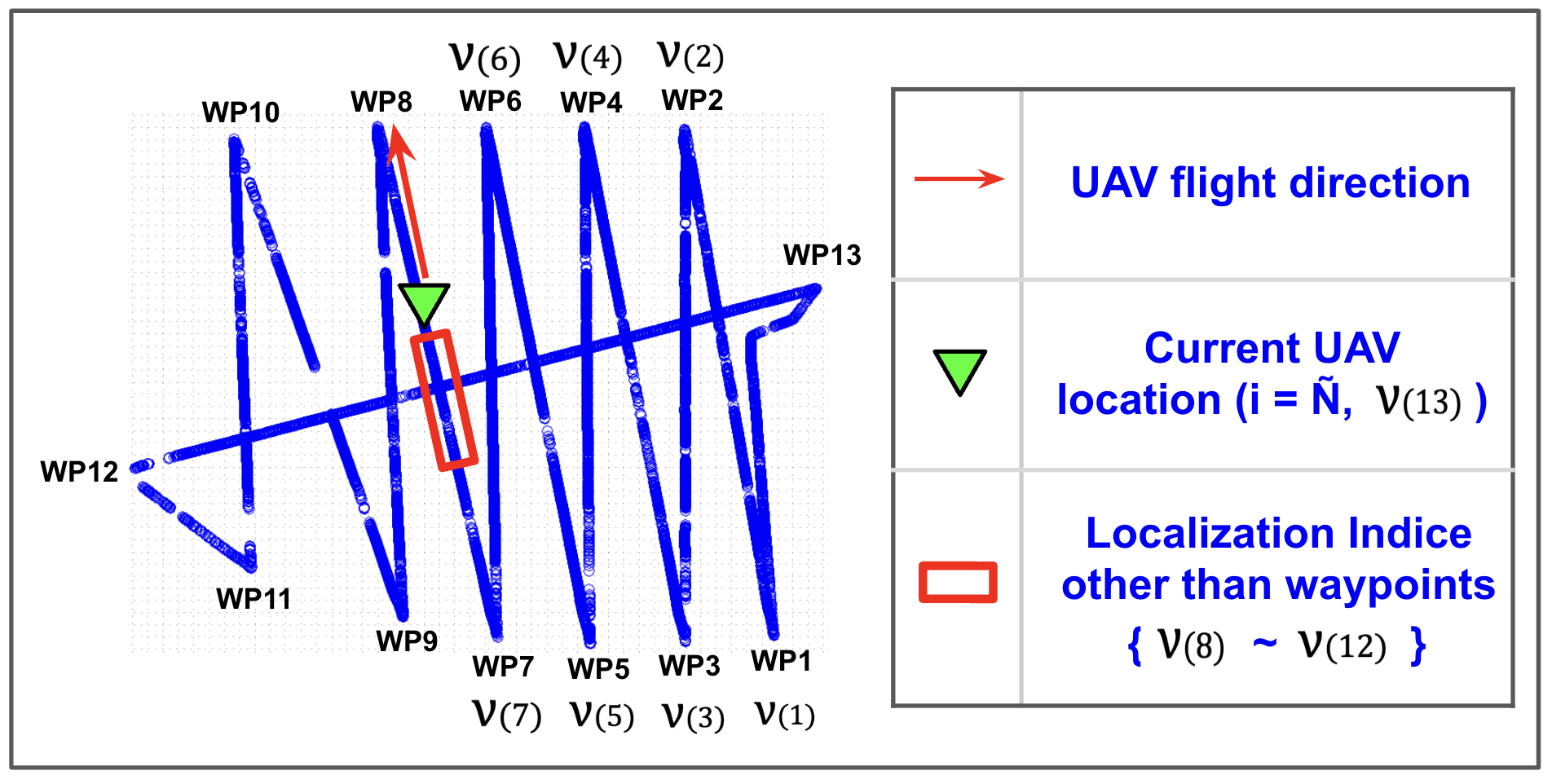}
\caption{LLS-CHLM}\label{fig:chlm}
%\vspace{-1mm}
\end{subfigure}
\begin{subfigure}{0.74\columnwidth}
\centering
\includegraphics[width=\textwidth]{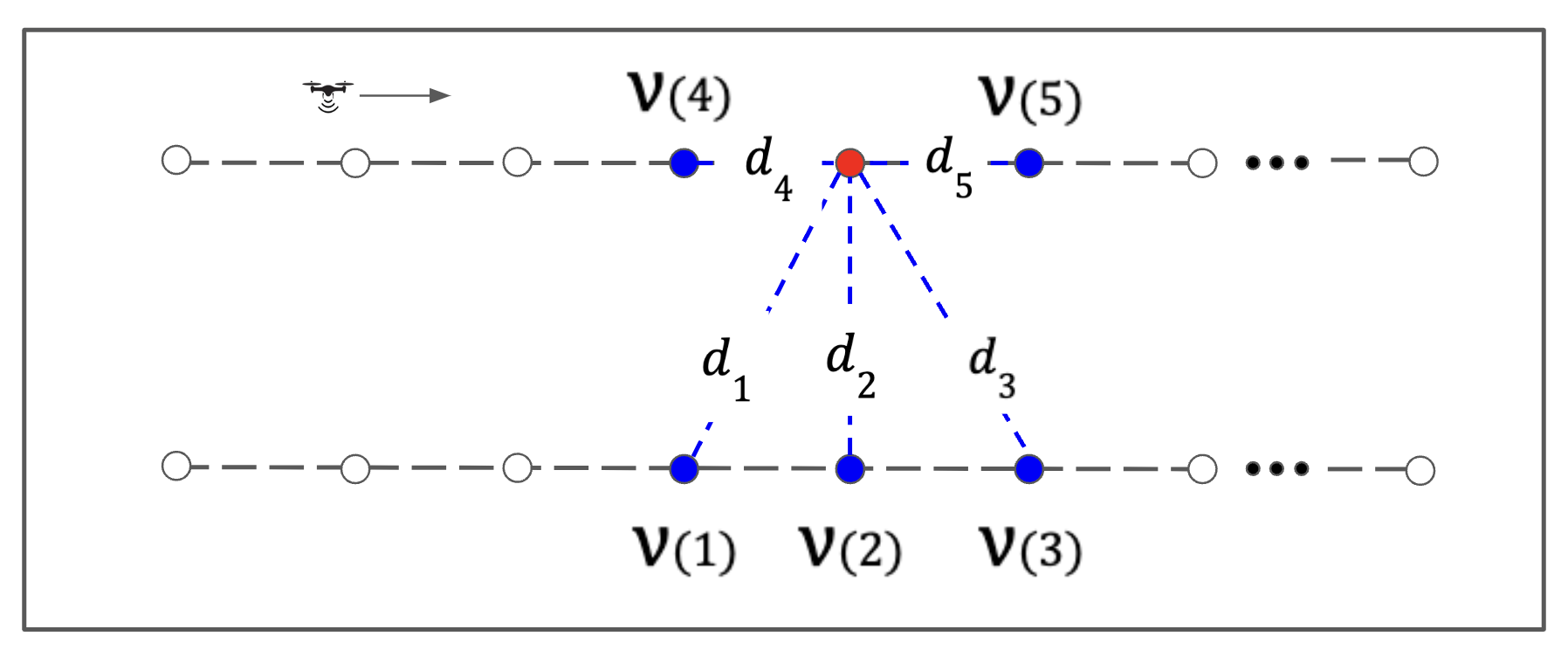}
\caption{LLS-CLS}\label{fig:cls}
%\vspace{-4mm}
\end{subfigure}
\caption{SSSL algorithms considered in this paper.}
\label{fig:algorithm}
\vspace{-3mm}
\end{figure}

In our previous study~\cite{kwon2023rf}, five heuristic-based localization algorithms were proposed for the selection of indices $\boldsymbol{\upsilon}$ on the UAV's trajectory for constructing equations as indicated in \eqref{eq:13}: 1) LLS-CON: utilizing three consecutive indices from the current UAV's flight index $\tilde{N}$, given by $\boldsymbol{\upsilon}\:= \big\{ \tilde{N}, \: \tilde{N}-1, \:\tilde{N}-2\big\}$; 2) LLS-CUM: using every index up to the current UAV's flight index $\tilde{N}$, denoted as $\boldsymbol{\upsilon}\:=
\big\{ 1, \:..., \:\tilde{N}\big\}$; 3) LLS-FML: selecting the first, middle, and last index, which can be expressed as $\boldsymbol{\upsilon}~=
\left\{ 1, \: {\rm round}\left(\frac{\tilde{N}+1}{2}\right), \:\tilde{N} \right\}$; 4) LLS-CHLM: adopting three to five indices using the concept of the convex hull in which the connected lines of each index encompass all other indices of the UAV's trajectory; and 5) LLS-CLS: selecting three indices that are the closest to a target location, which can be represented as $\boldsymbol{\upsilon}\:=
\big\{ \boldsymbol{\upsilon}_{\phi}(1), \: \boldsymbol{\upsilon}_{\phi}(2), \: \boldsymbol{\upsilon}_{\phi}(3) \big\}~,\nonumber$ where $\boldsymbol{\upsilon}_{\phi}$ is a reordered by ascending manner. 

In previous work, these algorithms were evaluated in a simplified simulation scenario, utilizing the free space path loss model and minimizing noise factors to reduce the inherent randomness of the simulations. On the other hand, the present study aims to investigate the performance of these localization algorithms in more realistic environments, considering the two-ray propagation model in an open field and using more realistic antenna patterns. In the current simulation setup, the interval between each index on the UAV's trajectory, where RF measurements from the target are taken, is much shorter than what is used in~\cite{kwon2023rf}. Hence, we decided not to implement LLS-CON. In addition, given its poor localization accuracy in~\cite{kwon2023rf}, LLS-FML is also not included in this study. 
In the end, the LLS-CUM, LLS-CHLM, and LLS-CLS are used and compared with each other in this paper. 

In the current research, the number of indices on the UAV's trajectory is updated as follows for different SSSL approaches to accommodate the new simulation environment. 
\subsubsection{LLS-CUM}
The same index array as the previous study is used as $\boldsymbol{\upsilon}\:=\big\{ 1, \:..., \:\tilde{N}\big\}$. 
\subsubsection{LLS-CHLM}
A total of thirteen indices are used based on the number of predefined trajectory's waypoints (WPs). In this modified algorithm, consecutive thirteen indices are used before the UAV reaches the WPs. However, as the UAV passes each WP, index array $\boldsymbol{\upsilon}$ is updated in a way to have a WP passing index as a member. The index array for the LLS-CHLM is denoted as
\begin{equation}%\label{eq:14}
\begin{split}
\boldsymbol{\upsilon}~=
\big\{\tilde{N} , \:\tilde{N}-1, ..., \:\tilde{N}- \left( 12-j\right), ..., \:{\rm WP}_{1}, ..., \:{\rm WP}_{j} \big\}~,\\
\; \tilde{N}\geq 13~, {\rm and} \; 2 \leq j \leq 13~,\nonumber
\end{split}
\end{equation}
where ${\rm WP}_{j}$ is the $j_{th}$ WP upon the preplanned trajectory.

\subsubsection{LLS-CLS}
Uses five trajectory indices for better performance. As the interval of each index becomes narrowed, the consecutive three points tend to have limited localization performance. As we assume that $\boldsymbol{\upsilon}_{\phi}$ is a reordered array of $\boldsymbol{\upsilon}$ in an ascending manner, the index array used for implementation of the LLS-CLS is represented as
\begin{equation}%\label{eq:15}
\boldsymbol{\upsilon}\:=
\big\{ \boldsymbol{\upsilon}_{\phi}(1), \: \boldsymbol{\upsilon}_{\phi}(2), \: \boldsymbol{\upsilon}_{\phi}(3), \boldsymbol{\upsilon}_{\phi}(4), \boldsymbol{\upsilon}_{\phi}(5) \big\}~.\nonumber
\end{equation}
Figure \ref{fig:algorithm} presents the underlying frameworks of each localization algorithm being considered in this study.

\section{Numerical Results}\label{Sec:NumResult}
\begin{figure}[t!]
\vspace{-1mm}
\centering
\includegraphics[trim=0.2cm 0.1cm 0.2cm 0.3cm, clip,width=9cm]{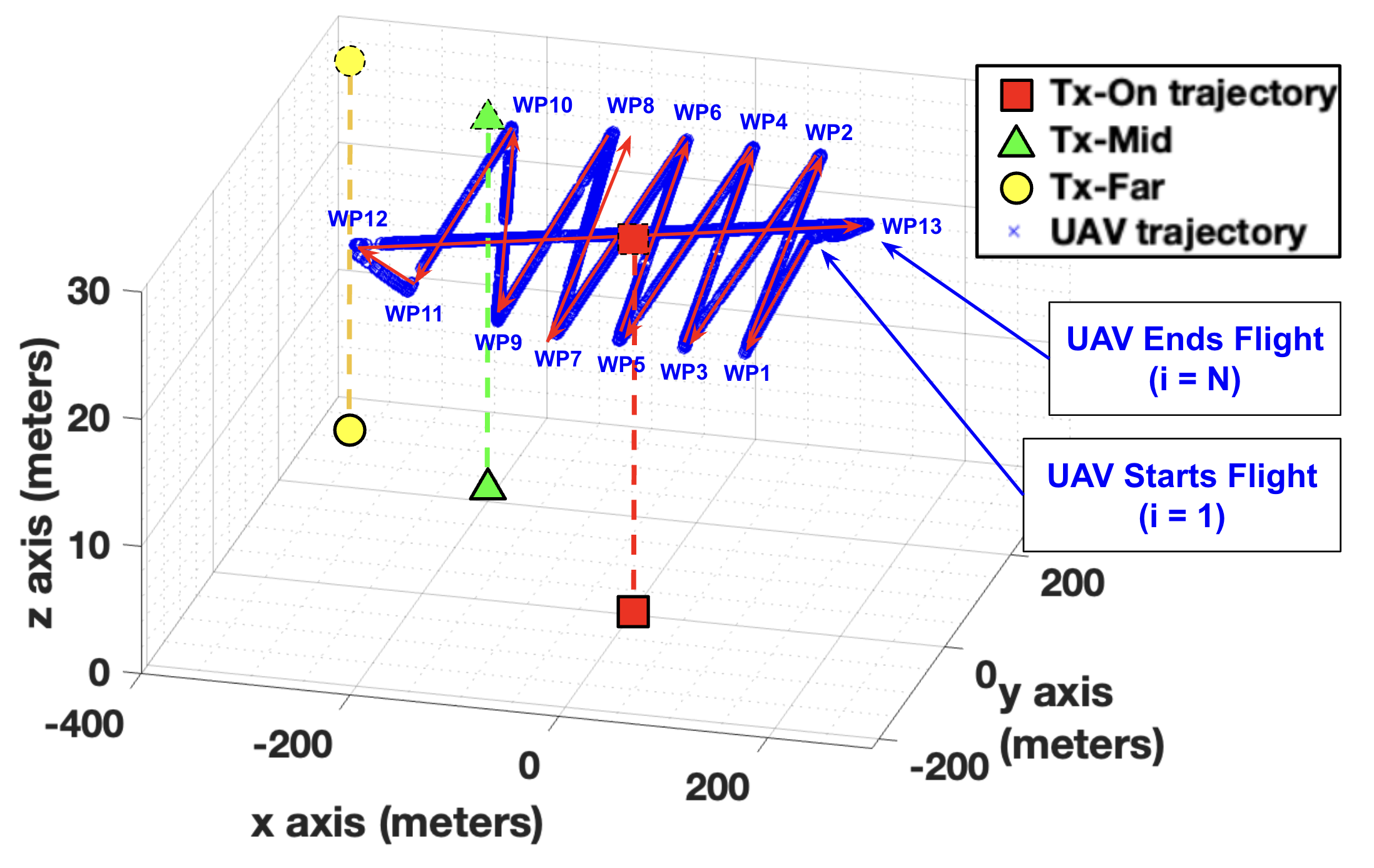}
\caption{Trajectory of the UAV with fixed waypoints, and three different signal source locations that the UAV is searching. While all the target locations are on the ground, their corresponding locations at the UAV's altitude are also illustrated, to show their relative location with respect to the UAV's trajectory more clearly.}
\label{fig:trajectory}
\vspace{-3mm}
\end{figure}

\begin{table}[t]\label{t1}
\caption{Summary of simulation parameters.}
\begin{center}
\scalebox{0.8}{
\begin{tabular}{p{4cm} c}
\hline
Channel model                   & \\ \hline
Carrier frequency               & 2.4 GHz                                 \\ 
Bandwidth                       & 20 MHz       \\ 
Path loss exponent              & 2          \\ \hline \hline
Transmitter                     & \\ \hline
Height                          & 1 m                                       \\ 
Transmit power                  & 76 dBm            \\ 
Antenna gain (Omni)             & 2 dBi      \\  \hline \hline
UAV                             & \\ \hline
Altitude                        & $\{30, 50, 70\}$ m \\
Antenna gain (Omni)             & 2 dBi  \\ \hline

\end{tabular}}
\end{center}
\vspace{-5mm}
\end{table}

% RSS figures
\begin{figure*}[!ht]
\centering
\begin{subfigure}{0.5\columnwidth}
\centering
\includegraphics[width=\textwidth]{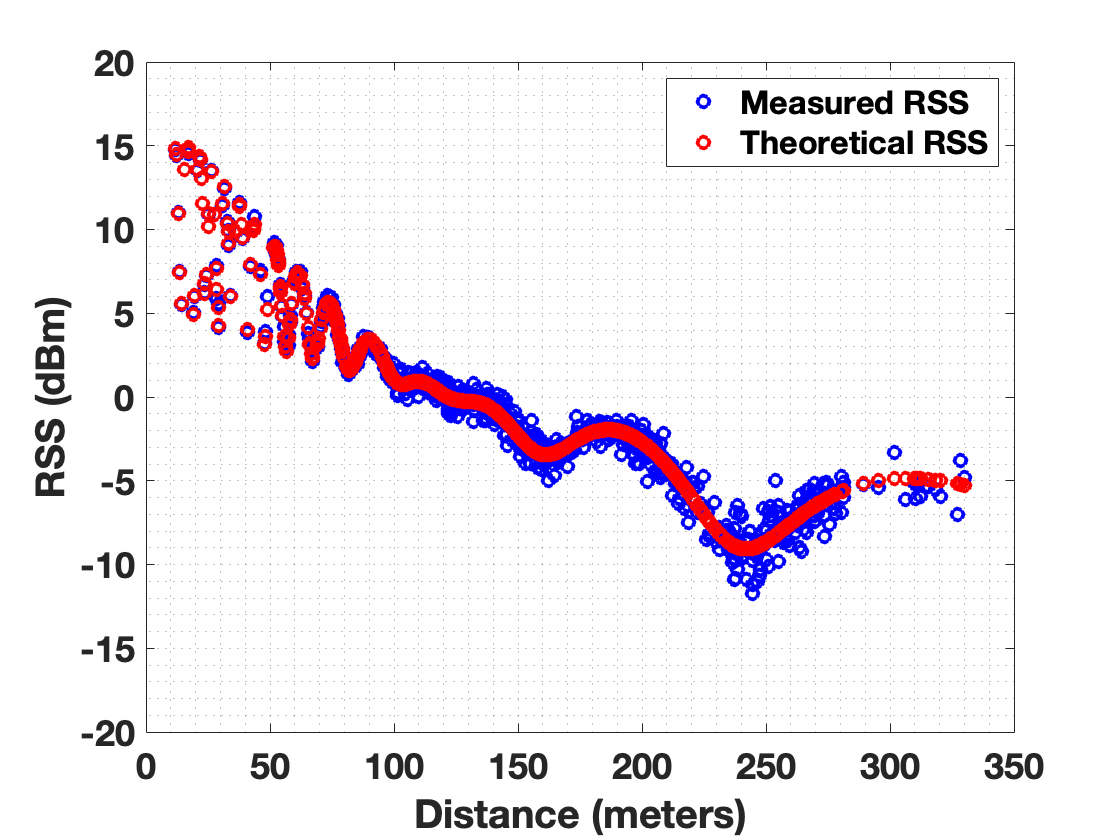}
\caption{Omnidirectional ($h=30$~m)}\label{fig:rssOmni}
%\vspace{-1mm}
\end{subfigure}
\begin{subfigure}{0.5\columnwidth}
\centering
\includegraphics[width=\textwidth]{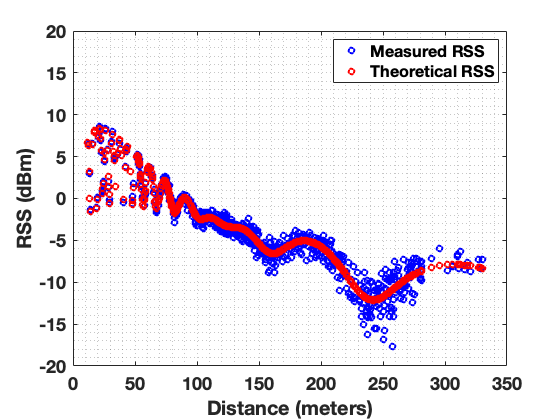}
\caption{Dipole antenna ($h=30$~m)}
\label{fig:rssDipole30}
%\vspace{-1mm}
\end{subfigure}
\begin{subfigure}{0.5\columnwidth}
\centering
\includegraphics[width=\textwidth]{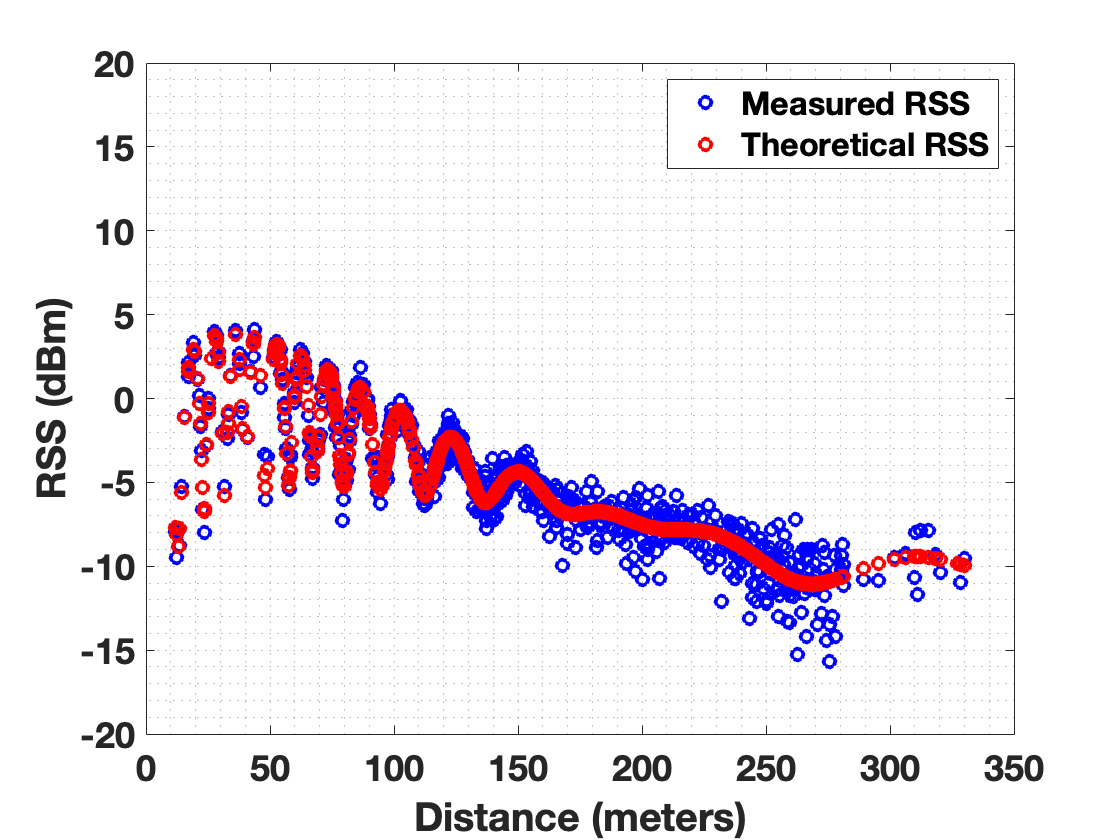}
\caption{Dipole antenna ($h=50$~m)}\label{fig:rssDipole50}
%\vspace{-1mm}
\end{subfigure}
\begin{subfigure}{0.5\columnwidth}
\centering
\includegraphics[width=\textwidth]{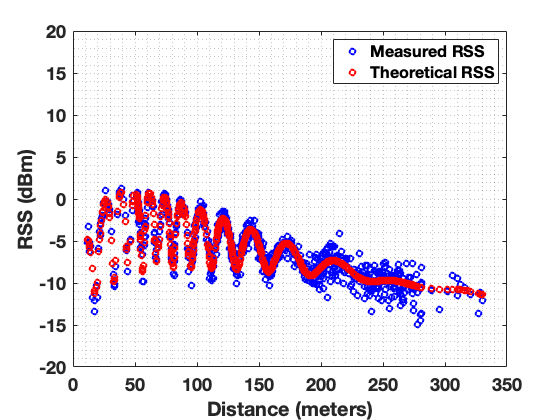}
\caption{Dipole antenna ($h=70$~m)}\label{fig:rssDipole70}
%\vspace{-1mm}
\end{subfigure}
\begin{subfigure}{0.5\columnwidth}
\centering
\includegraphics[width=\textwidth]{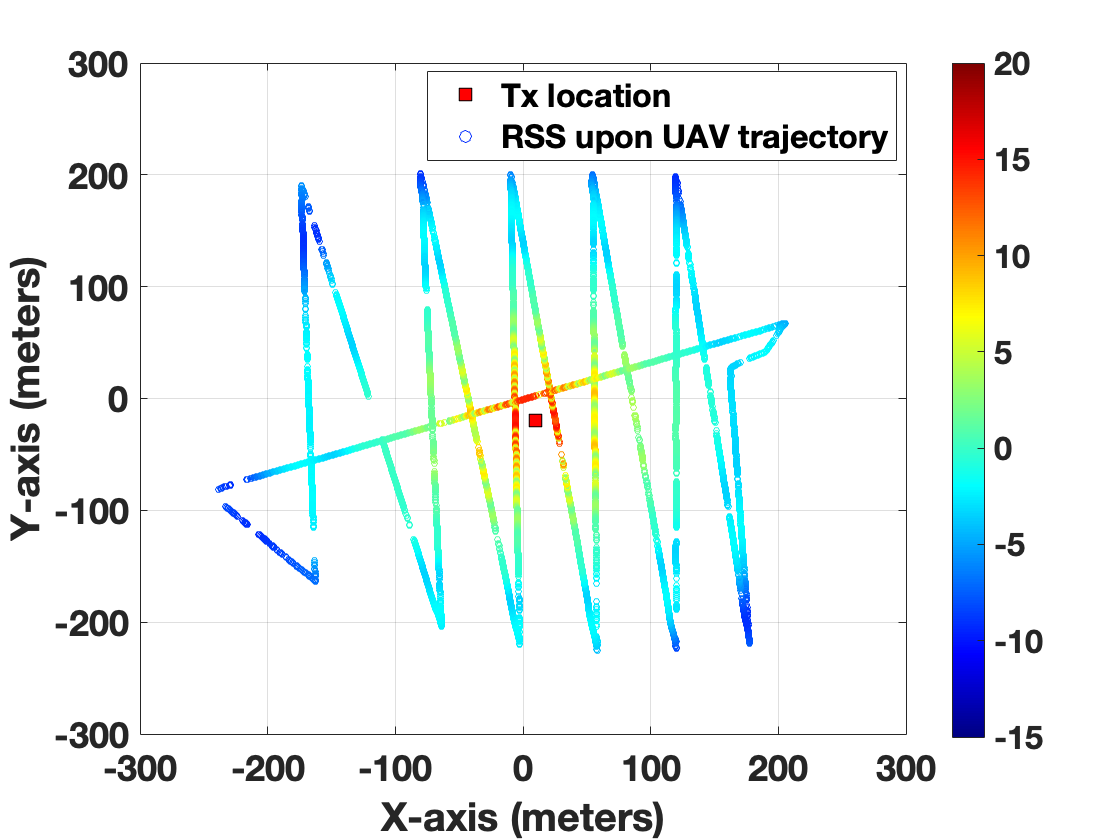}
\caption{Omnidirectional ($h=30$~m)}
\label{fig:rssTrajOmni}
%\vspace{-1mm}
\end{subfigure}
\begin{subfigure}{0.5\columnwidth}
\centering
\includegraphics[width=\textwidth]{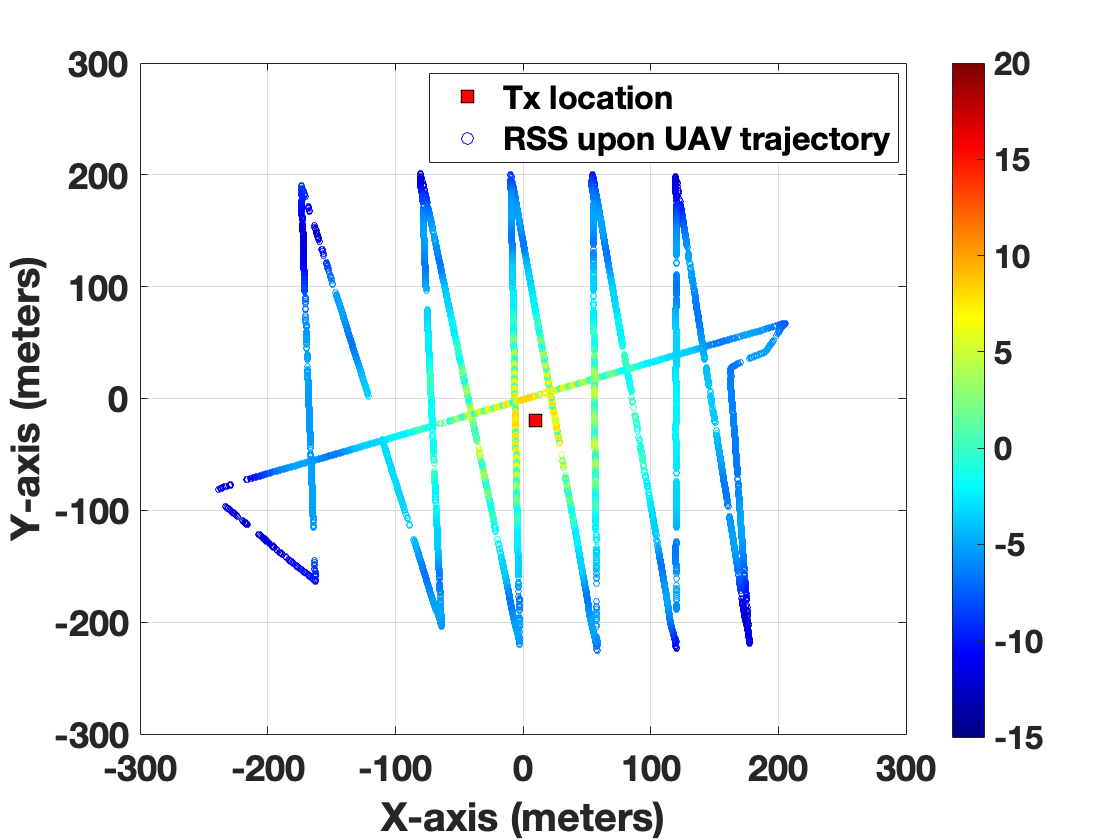}
\caption{Dipole antenna ($h=30$~m)}\label{fig:rssTrajDipole30}
%\vspace{-1mm}
\end{subfigure}
\begin{subfigure}{0.5\columnwidth}
\centering
\includegraphics[width=\textwidth]{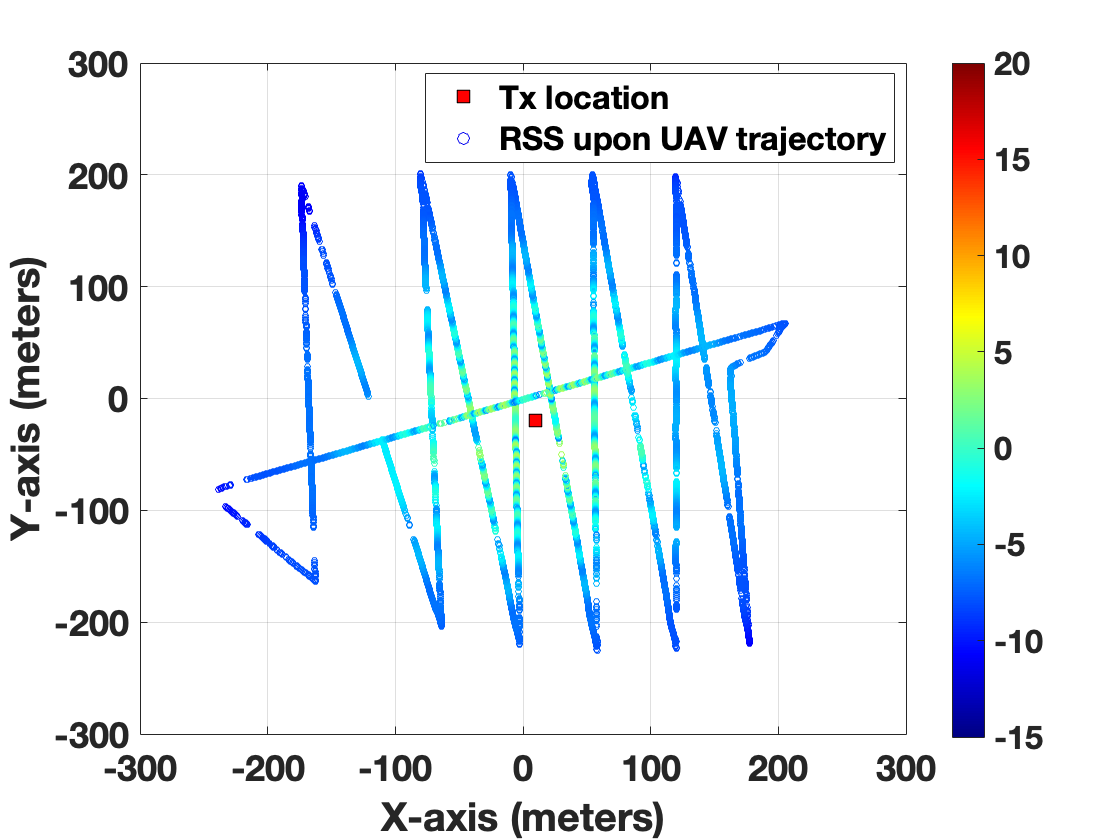}
\caption{Dipole antenna ($h=50$~m)}
\label{fig:rssTrajDipole60}
%\vspace{-1mm}
\end{subfigure}
\begin{subfigure}{0.5\columnwidth}
\centering
\includegraphics[width=\textwidth]{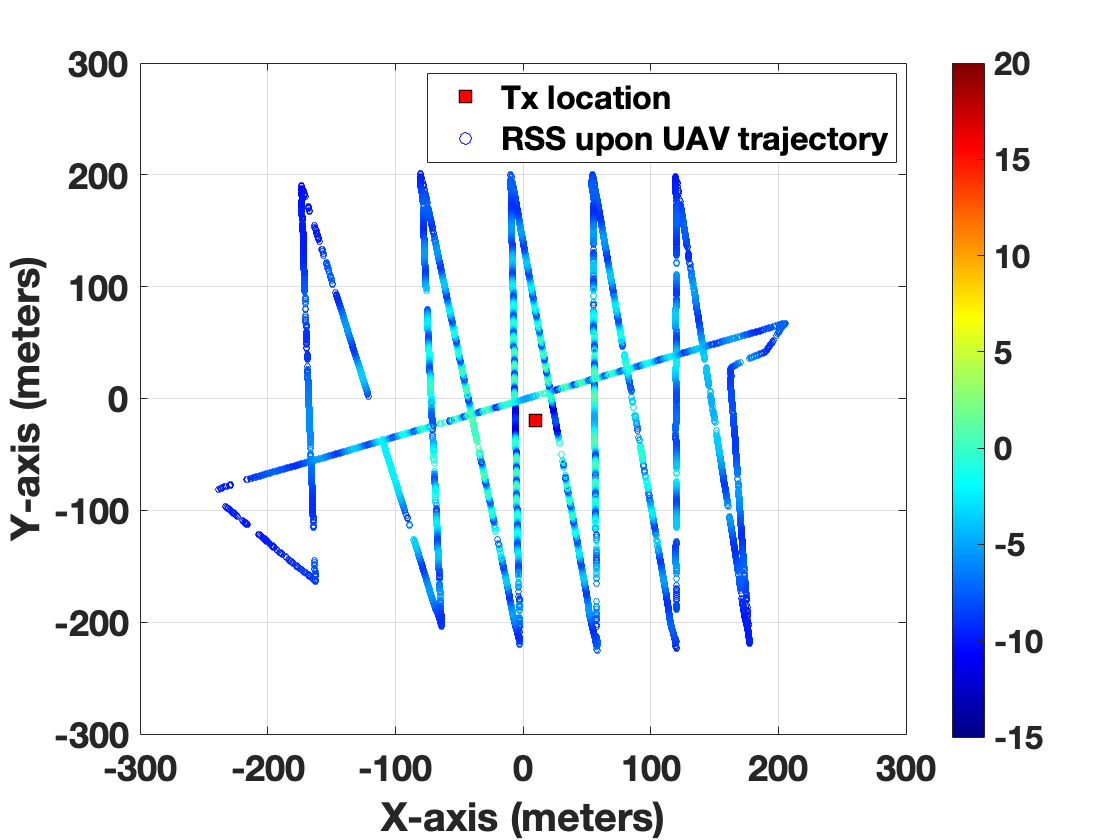}
\caption{Dipole antenna ($h=70$~m)}\label{fig:rssTrajDipole70}
%\vspace{-1mm}
\end{subfigure}
\caption{RSS measurements for various scenarios. (a)-(d): RSS vs. distance, and (e)-(h) RSS vs. trajectory.}
\label{fig:simulMeasure}
\end{figure*}

% Distance and angle relations
\begin{figure*}[!ht]
\centering
\begin{subfigure}{0.5\columnwidth}
\centering
\includegraphics[width=\textwidth]{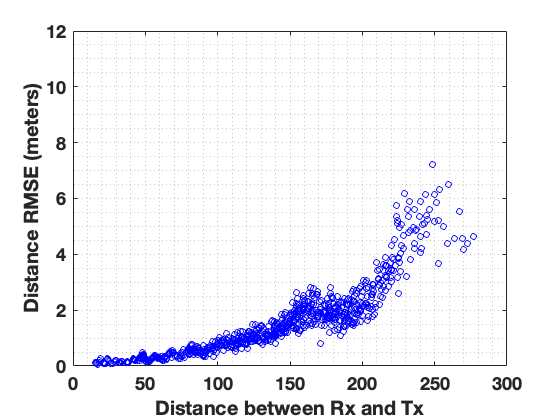}
\caption{DROD: Omni ($h=30$~m)}\label{fig:distrmseOmni}
%\vspace{-1mm}
\end{subfigure}
\begin{subfigure}{0.5\columnwidth}
\centering
\includegraphics[width=\textwidth]{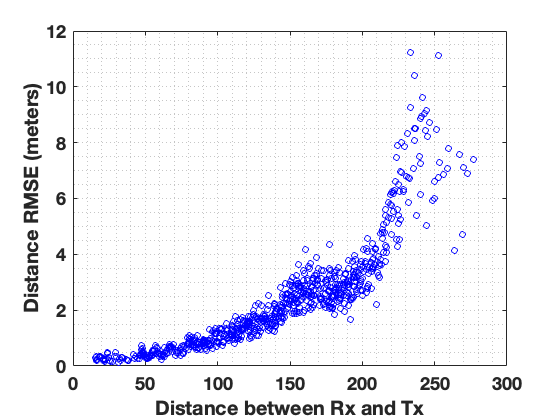}
\caption{DROD: Dipole ($h=30$~m)}\label{fig:distDistRmseDipole30}
%\vspace{-1mm}
\end{subfigure}
\begin{subfigure}{0.5\columnwidth}
\centering
\includegraphics[width=\textwidth]{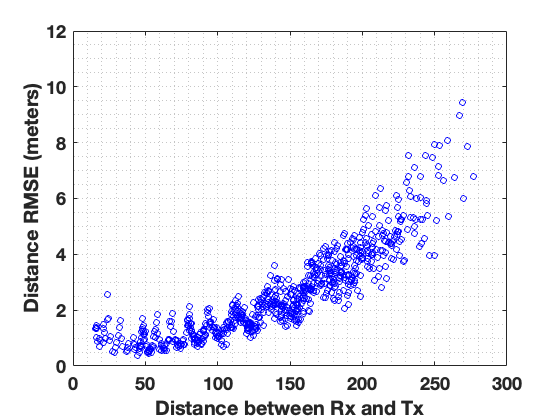}
\caption{DROD: Dipole ($h=50$~m)}\label{fig:distDistRmseDipole50}
%\vspace{-1mm}
\end{subfigure}
\begin{subfigure}{0.5\columnwidth}
\centering
\includegraphics[width=\textwidth]{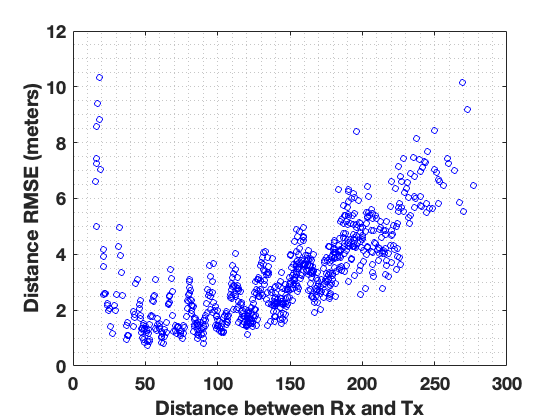}
\caption{DROD: Dipole ($h=70$~m)}\label{fig:distDistRmseDipole70}
%\vspace{-1mm}
\end{subfigure}

\begin{subfigure}{0.5\columnwidth}
\centering
\includegraphics[width=\textwidth]{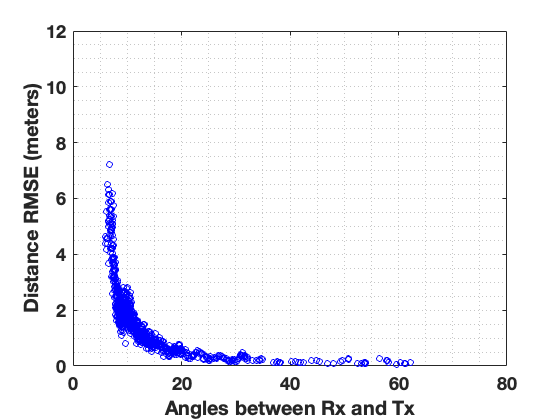}
\caption{DROA: Omni ($h=30$~m)}\label{fig:angDistRmseOmni}
%\vspace{-1mm}
\end{subfigure}
\begin{subfigure}{0.5\columnwidth}
\centering
\includegraphics[width=\textwidth]{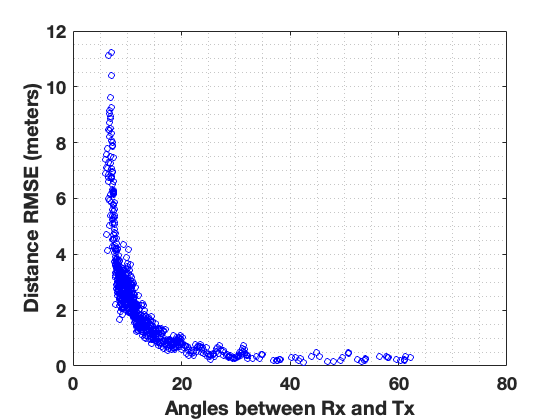}
\caption{DROA: Dipole ($h=30$~m)}\label{fig:angDistRmseDipole30}
%\vspace{-1mm}
\end{subfigure}
\begin{subfigure}{0.5\columnwidth}
\centering
\includegraphics[width=\textwidth]{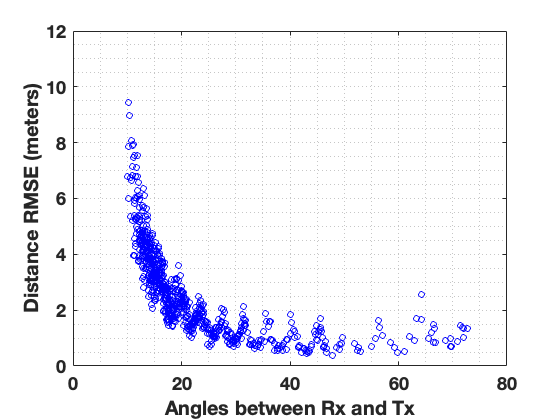}
\caption{DROA: Dipole ($h=50$~m)}\label{fig:angDistRmseDipole50}
%\vspace{-1mm}
\end{subfigure}
\begin{subfigure}{0.5\columnwidth}
\centering
\includegraphics[width=\textwidth]{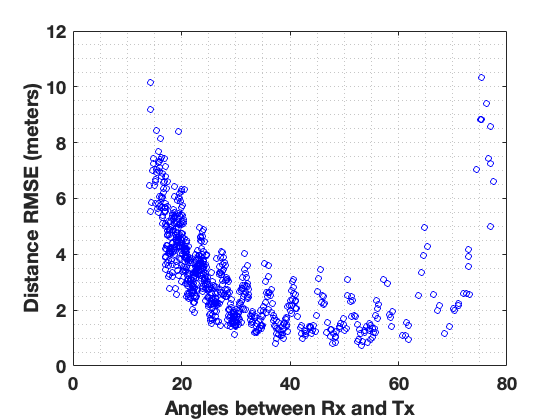}
\caption{DROA: Dipole ($h=70$~m)}\label{fig:angDistRmseDipole70}
%\vspace{-1mm}
\end{subfigure}

\begin{subfigure}{0.5\columnwidth}
\centering
\includegraphics[width=\textwidth]{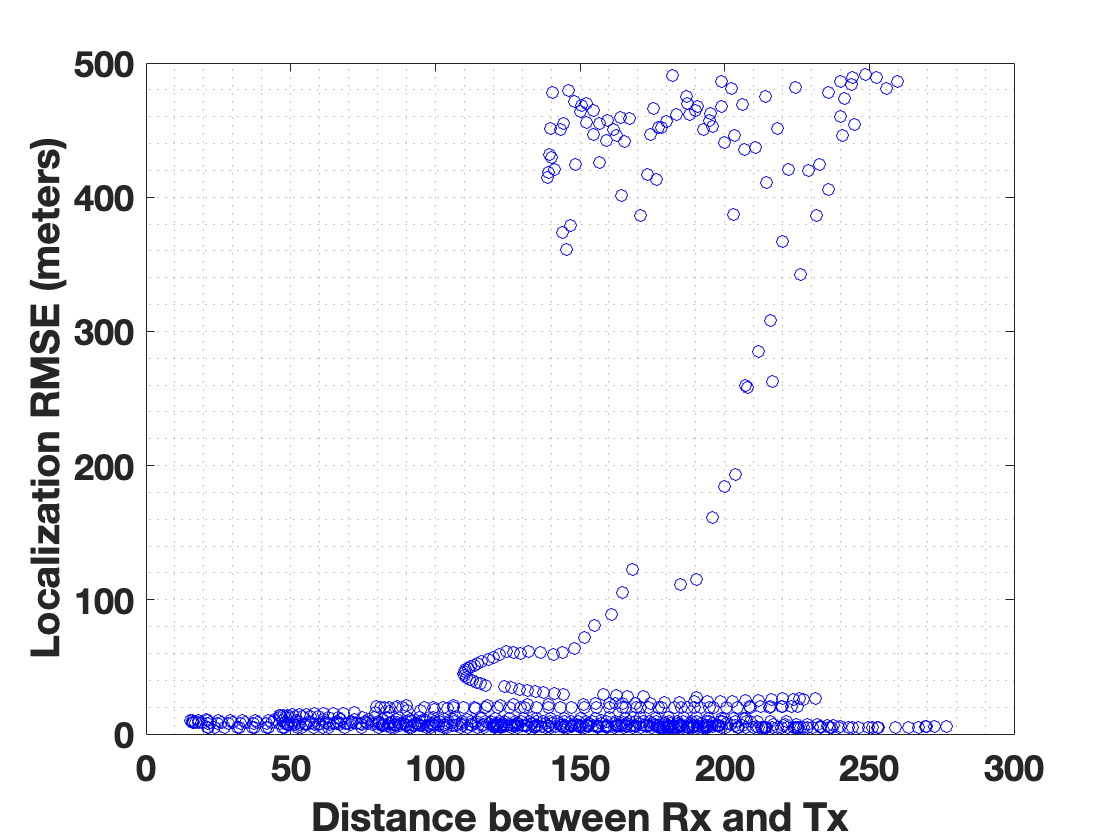}
\caption{LROD: Omni ($h=30$~m)}\label{fig:distLocalRmseOmni}
%\vspace{-1mm}
\end{subfigure}
\begin{subfigure}{0.5\columnwidth}
\centering
\includegraphics[width=\textwidth]{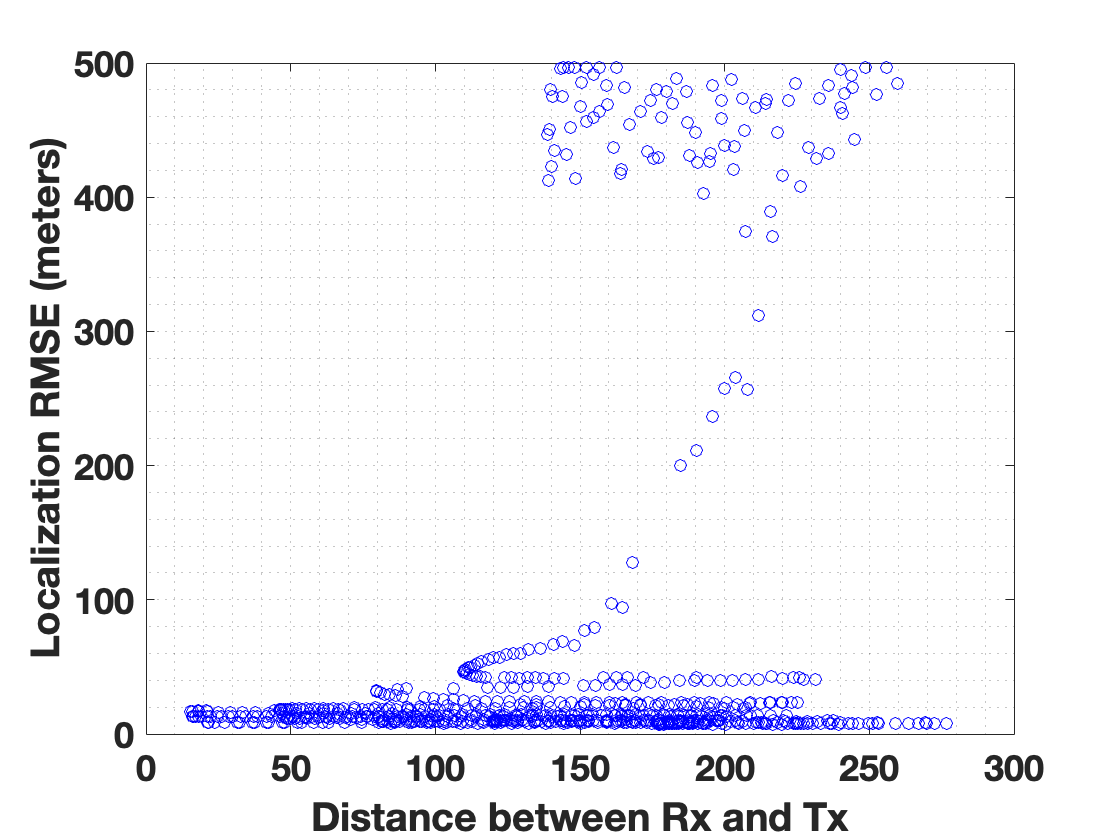}
\caption{LROD: Dipole ($h=30$~m)}\label{fig:distLocalRmseDipole30}
%\vspace{-1mm}
\end{subfigure}
\begin{subfigure}{0.5\columnwidth}
\centering
\includegraphics[width=\textwidth]{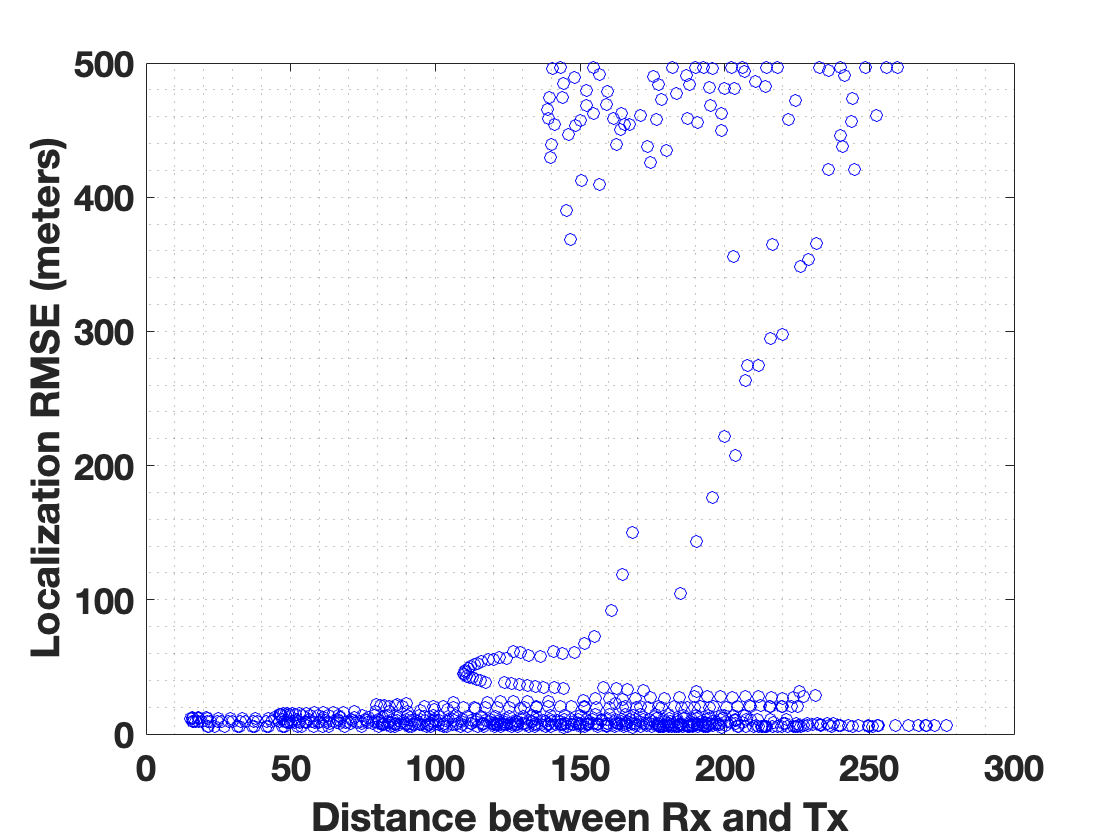}
\caption{LROD: Dipole ($h=50$~m)}\label{fig:distLocalRmseDipole50}
%\vspace{-1mm}
\end{subfigure}
\begin{subfigure}{0.5\columnwidth}
\centering
\includegraphics[width=\textwidth]{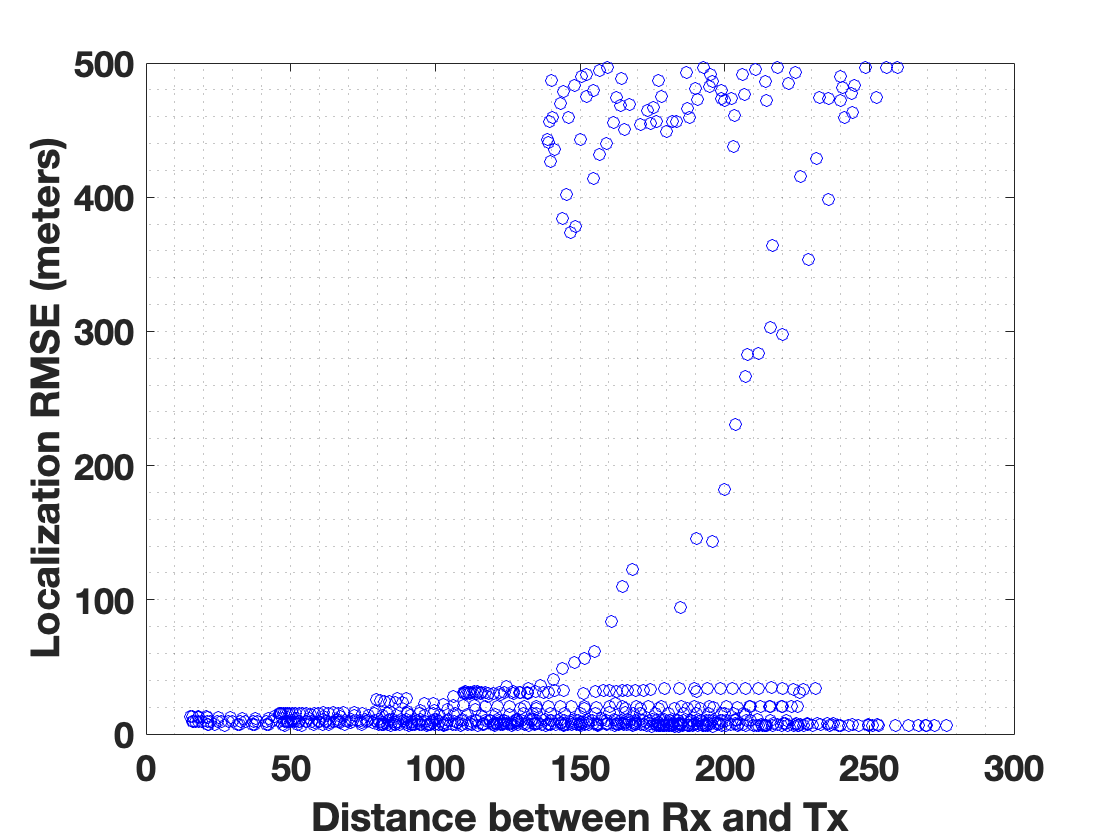}
\caption{LROD: Dipole ($h=70$~m)}\label{fig:distLocalRmseDipole70}
%\vspace{-1mm}
\end{subfigure}

\begin{subfigure}{0.5\columnwidth}
\centering
\includegraphics[width=\textwidth]{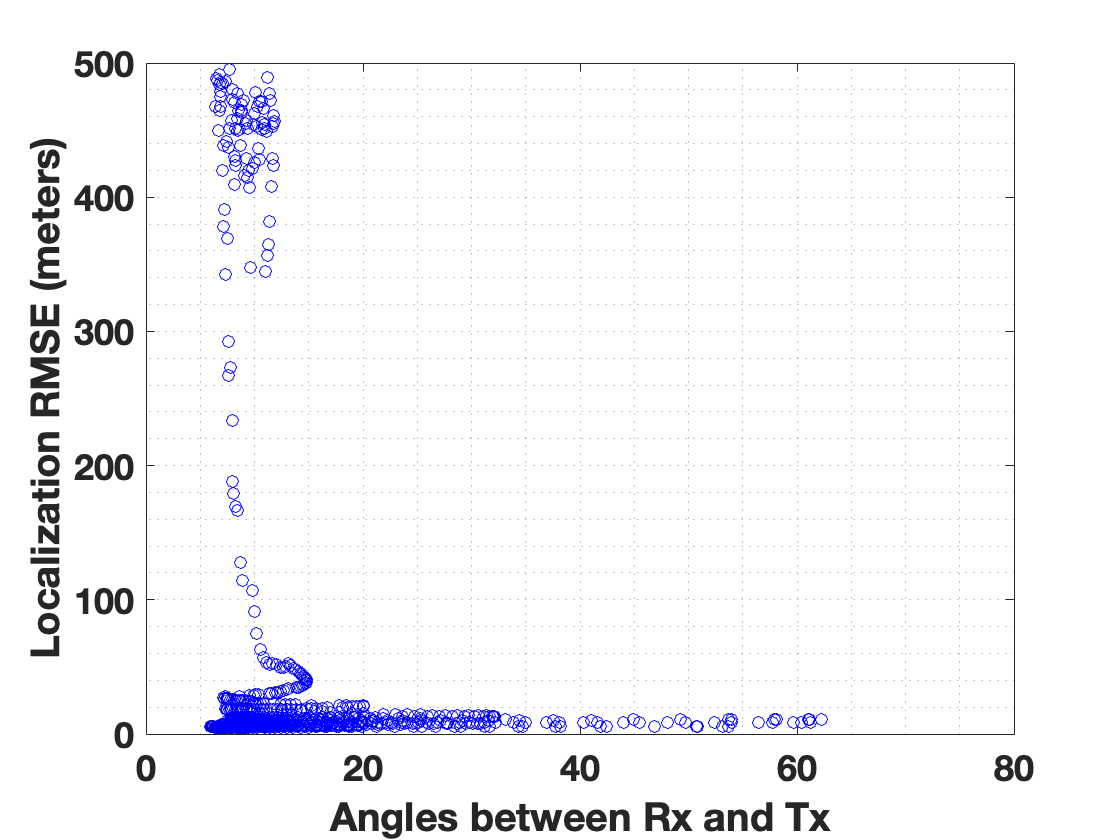}
\caption{LROA: Omni ($h=30$~m)}\label{fig:angLocalRmseOmni}
%\vspace{-1mm}
\end{subfigure}
\begin{subfigure}{0.5\columnwidth}
\centering
\includegraphics[width=\textwidth]{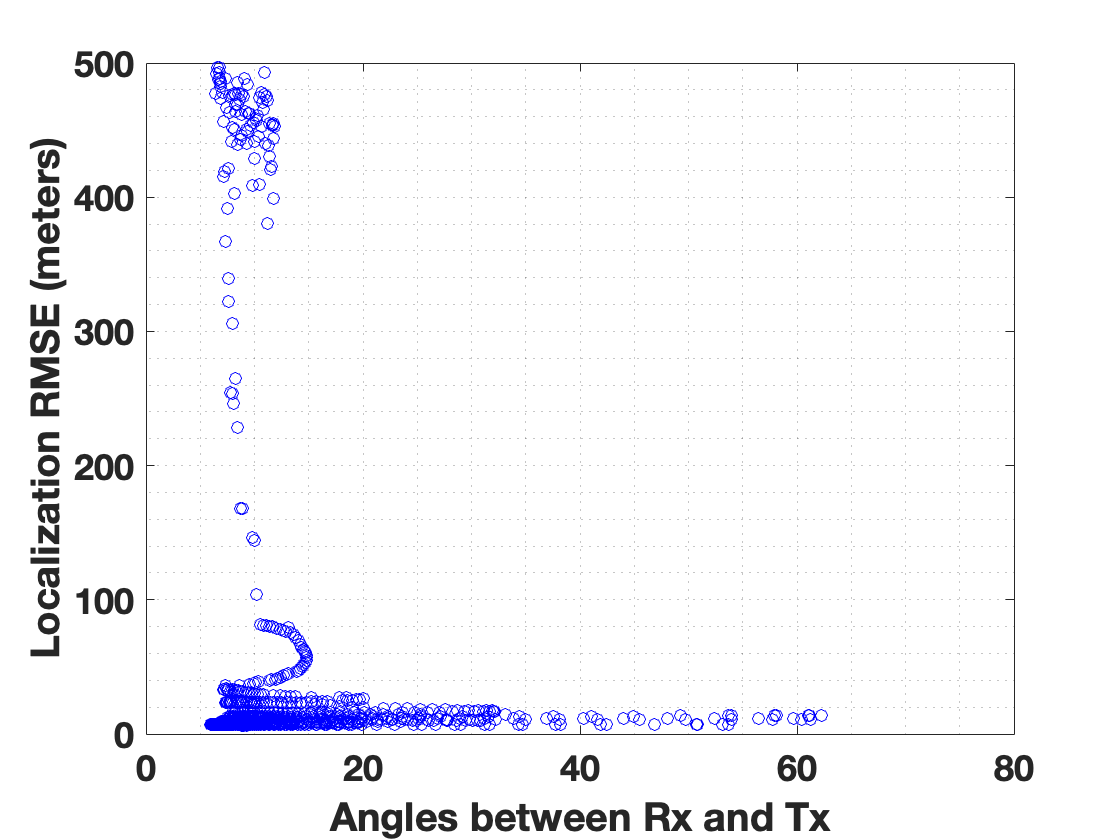}
\caption{LROA: Dipole ($h=30$~m)}\label{fig:angLocalRmseDipole30}
%\vspace{-1mm}
\end{subfigure}
\begin{subfigure}{0.5\columnwidth}
\centering
\includegraphics[width=\textwidth]{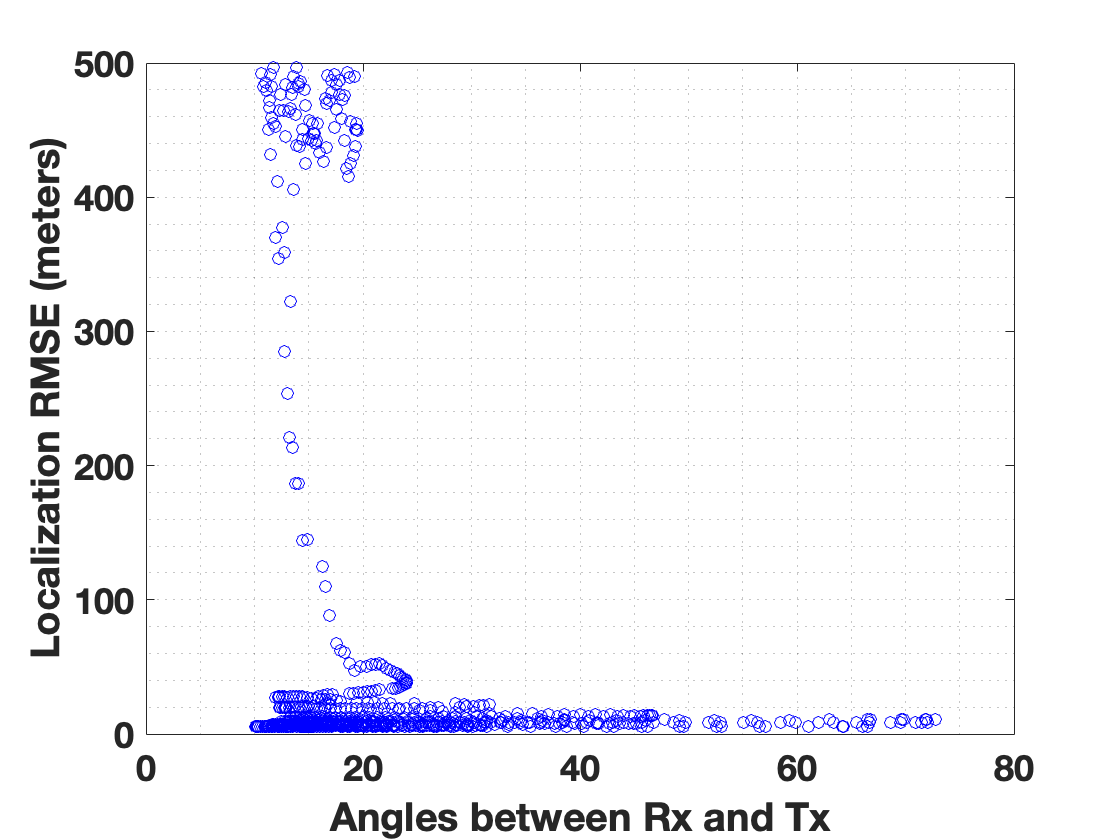}
\caption{LROA: Dipole ($h=50$~m)}\label{fig:angLocalRmseDipole50}
%\vspace{-1mm}
\end{subfigure}
\begin{subfigure}{0.5\columnwidth}
\centering
\includegraphics[width=\textwidth]{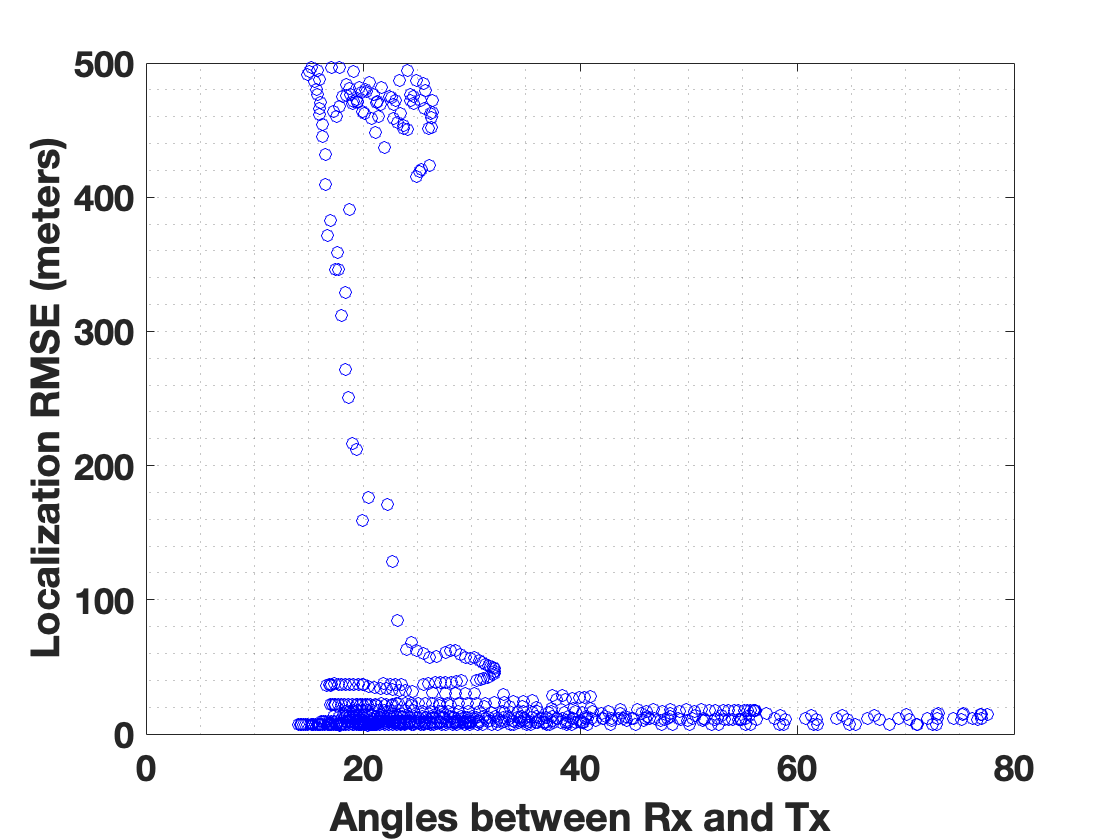}
\caption{LROA: Dipole ($h=70$~m)}\label{fig:angLocalRmseDipole70}
%\vspace{-1mm}
\end{subfigure}
\caption{Estimation RMSE for distance and angle variation. (a)-(h): distance estimation, and (i)-(p): localization.}
\vspace{-1.5mm}
\label{fig:rmselMeasure}
\end{figure*}

The experiments are carried out in MATLAB-based simulations. During the simulation stage, the theoretical received signal strength (RSS) is evaluated based on the two-ray propagation model. The three proposed localization algorithms are then implemented and comparatively assessed in terms of their localization performance. In this stage, the transmitter's and receiver's antenna patterns are set to two separate configurations: 1) omnidirectional antenna pattern; and 2)  doughnut-shaped dipole antenna pattern. Finally, to evaluate localization algorithm performance, three components were implemented consistent with the previous study: 1) accuracy; 2) time used to achieve a specific level of localization accuracy; and 3) reliability.

In simulations, the frequency and bandwidth for the transmitted signal are set as 2.4~GHz and 20~MHz, respectively. We employed a zig-zag pattern for the UAV's trajectory. In addition, the transmitter is positioned in three different locations: 1) the target is on the UAV's predefined trajectory; 2) the target is at the boundary of the trajectory; and 3) the target is in a position far away from the trajectory. This approach enabled the analysis of localization performance based on the distance of the signal source from the predefined trajectory. The trajectory and the different target locations are represented in Fig.~\ref{fig:trajectory}. For the antenna pattern configuration, the omnidirectional antenna pattern's gain is set to 2~dBi for simplicity. On the other hand, the doughnut-shaped dipole antenna gain is set based on \eqref{eq:2}. The principal distinction between the dipole antenna and the omnidirectional antenna pattern lies in the variation of antenna gain as the UAV's altitude changes.

% Localization algorithm comparison
\begin{figure*}[!ht]
\centering
\begin{subfigure}{0.63\columnwidth}
\centering
\includegraphics[width=\textwidth]{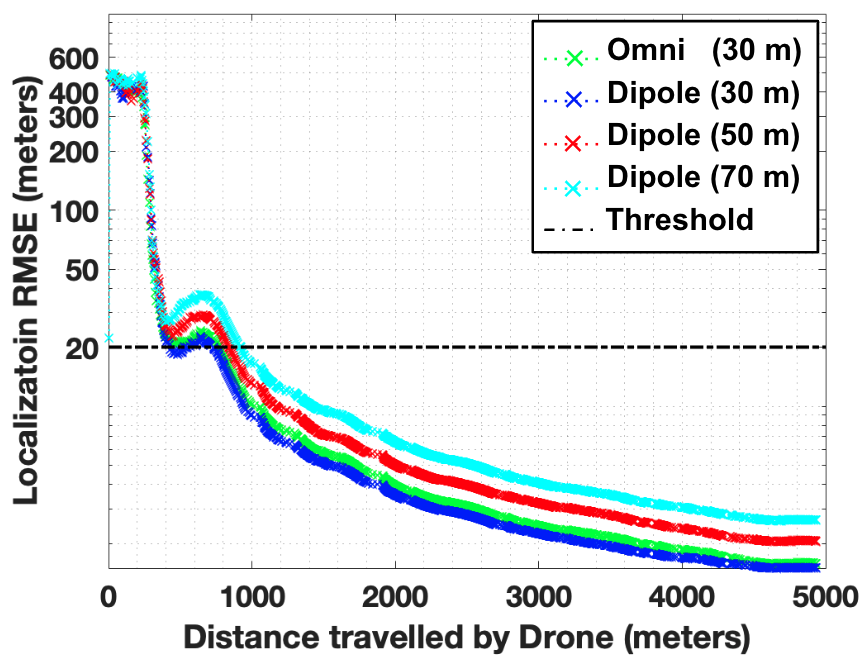}
\caption{On target: LLS-CUM}\label{fig:onCum}
%\vspace{-1mm}
\end{subfigure}
\begin{subfigure}{0.63\columnwidth}
\centering
\includegraphics[width=\textwidth]{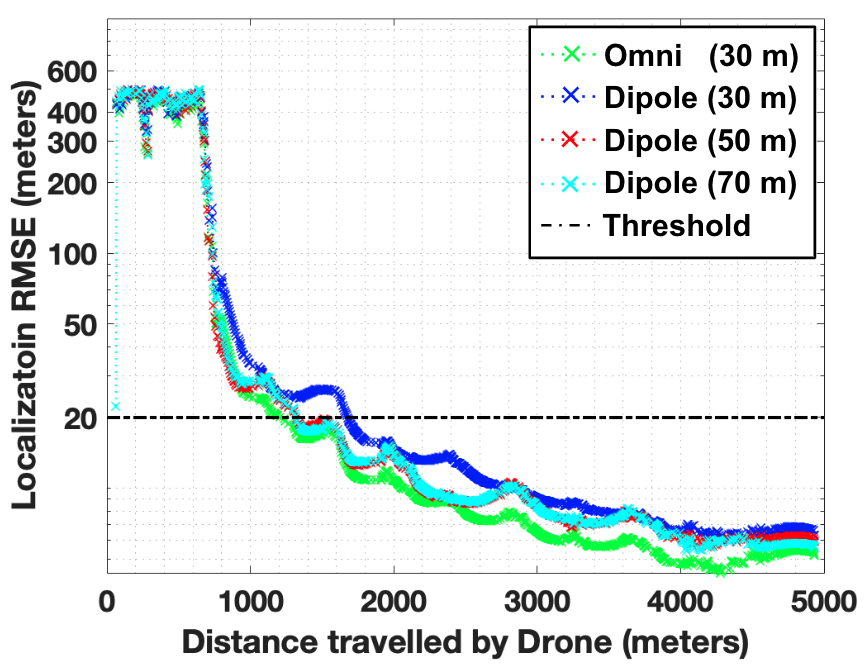}
\caption{On target: LLS-CHLM}\label{fig:onChml}
%\vspace{-1mm}
\end{subfigure}
\begin{subfigure}{0.63\columnwidth}
\centering
\includegraphics[width=\textwidth]{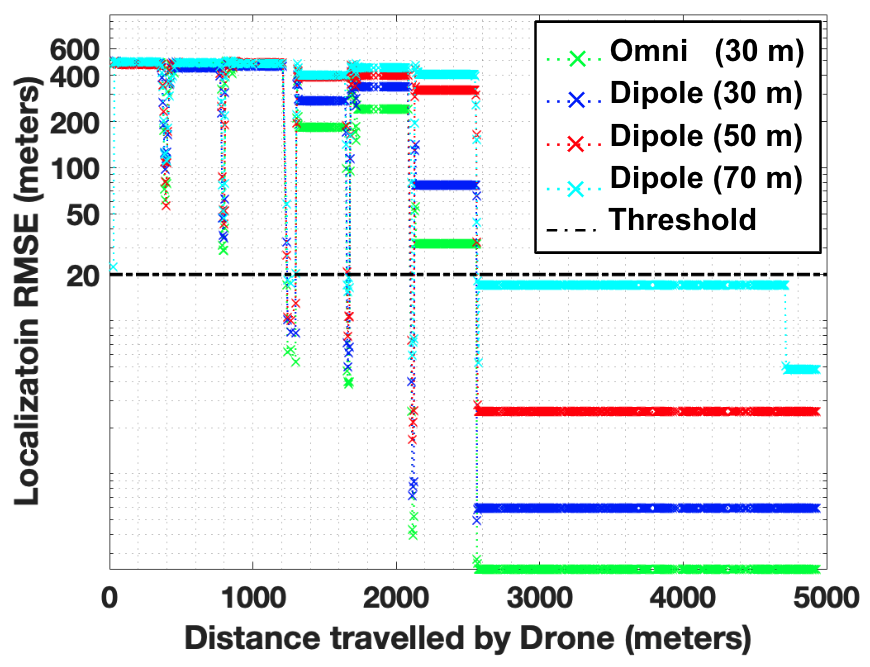}
\caption{On target: LLS-CLS}\label{fig:onCls}
%\vspace{-1mm}
\end{subfigure}

\begin{subfigure}{0.63\columnwidth}
\centering
\includegraphics[width=\textwidth]{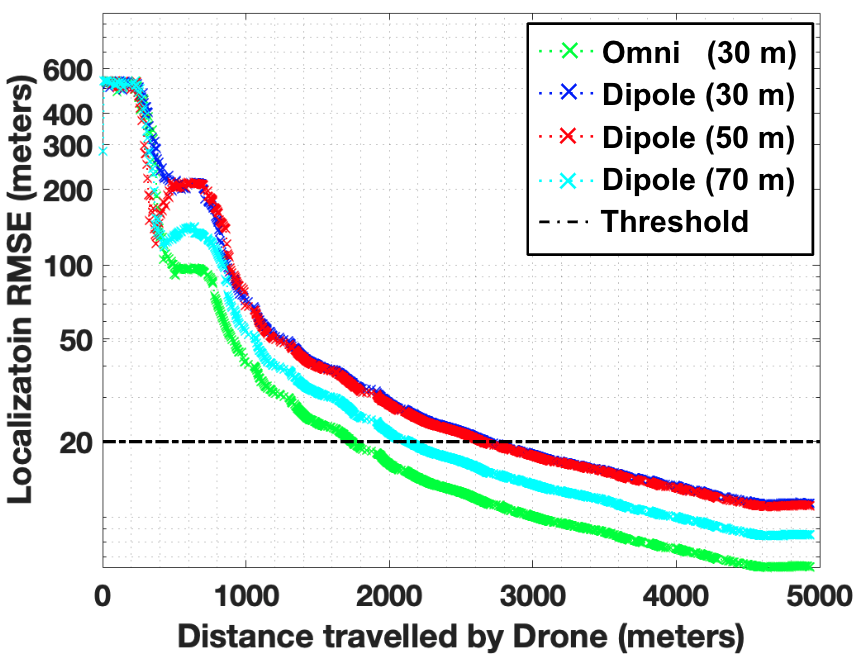}
\caption{Mid target: LLS-CUM}\label{fig:midCum}
%\vspace{-1mm}
\end{subfigure}
\begin{subfigure}{0.63\columnwidth}
\centering
\includegraphics[width=\textwidth]{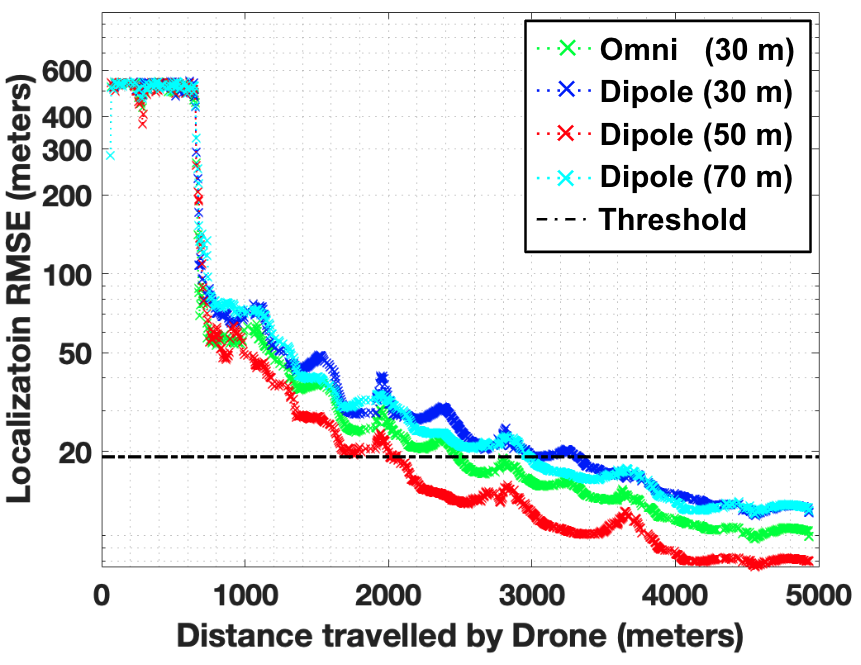}
\caption{Mid target: LLS-CHLM}\label{fig:midChml}
%\vspace{-1mm}
\end{subfigure}
\begin{subfigure}{0.63\columnwidth}
\centering
\includegraphics[width=\textwidth]{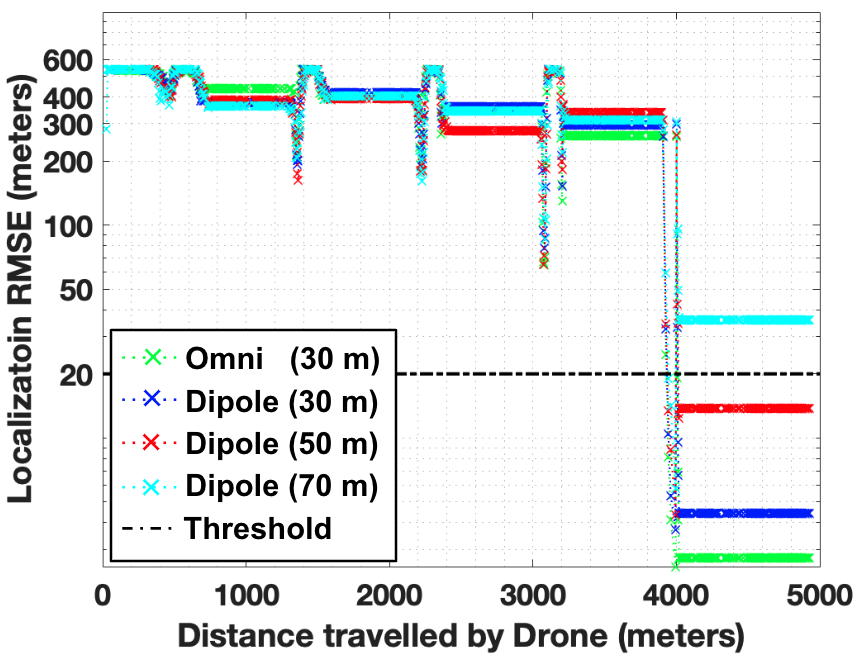}
\caption{Mid target: LLS-CLS}\label{fig:midCls}
%\vspace{-1mm}
\end{subfigure}

\begin{subfigure}{0.63\columnwidth}
\centering
\includegraphics[width=\textwidth]{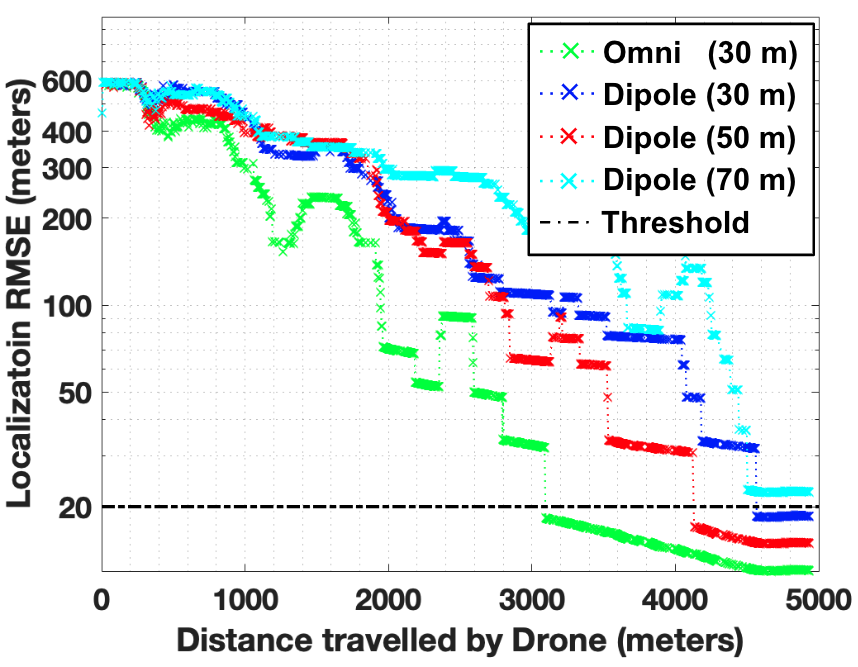}
\caption{Far target: LLS-CUM}\label{fig:farCum}
%\vspace{-1mm}
\end{subfigure}
\begin{subfigure}{0.63\columnwidth}
\centering
\includegraphics[width=\textwidth]{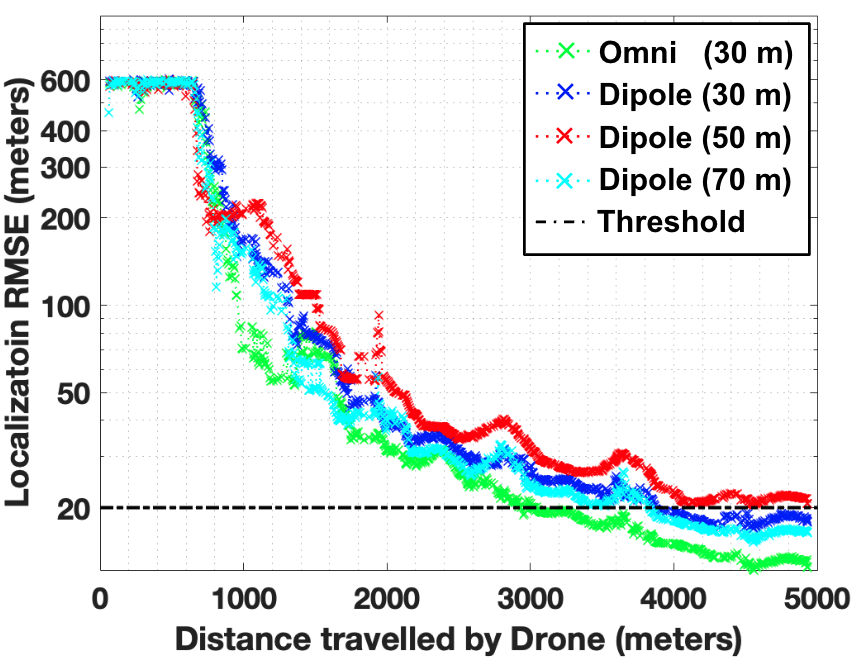}
\caption{Far target: LLS-CHLM}\label{fig:farChml}
%\vspace{-1mm}
\end{subfigure}
\begin{subfigure}{0.63\columnwidth}
\centering
\includegraphics[width=\textwidth]{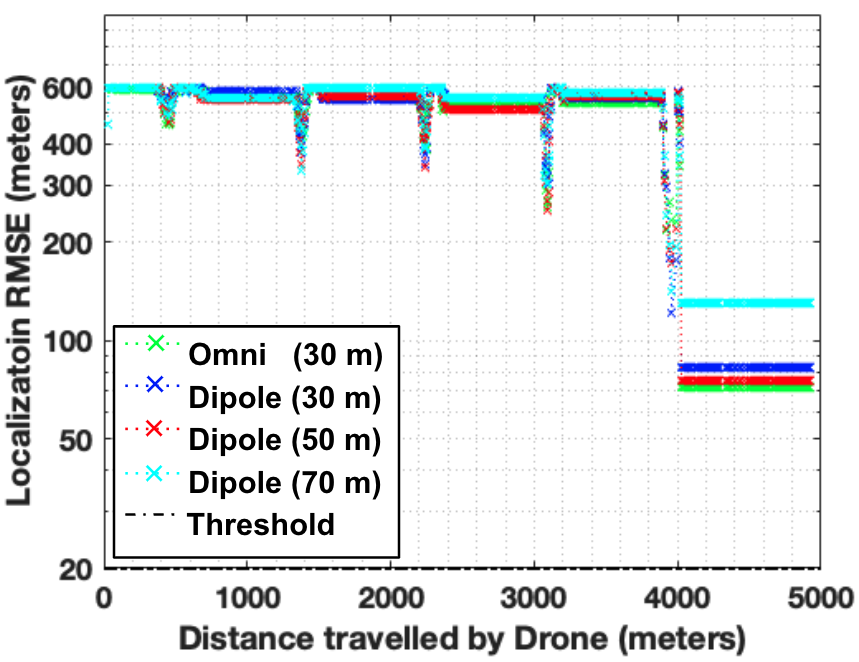}
\caption{Far target: LLS-CLS}\label{fig:farCls}
%\vspace{-1mm}
\end{subfigure}
\caption{Localization RMSE comparison upon various target locations: Omni, Dipole (with altitude 30m, 50m, and 70m)}\vspace{-3mm}
\label{fig:compLocal}
\end{figure*}

% Localization performance upon target location
\begin{figure*}[!ht]
\centering
\begin{subfigure}{0.95\columnwidth}
\centering
\includegraphics[width=\textwidth]{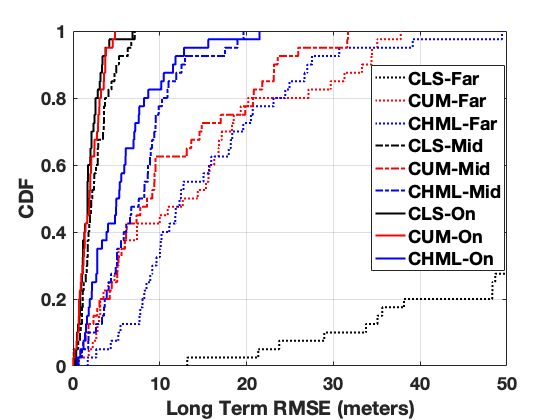}
\caption{Long term RMSE CDF}\label{fig:localcdf}
%\vspace{-1mm}
\end{subfigure}
\begin{subfigure}{0.95\columnwidth}
\centering
\includegraphics[width=\textwidth]{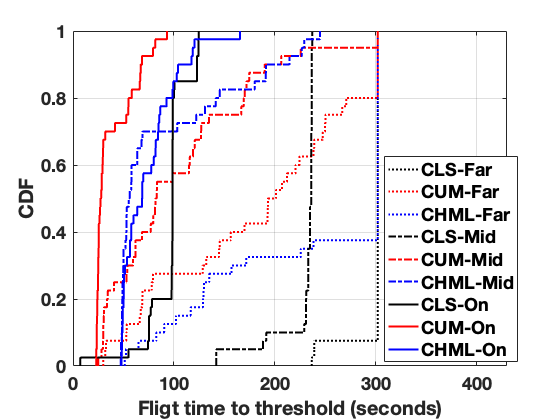}
\caption{Flight time CDF}\label{fig:timecdf}
%\vspace{-1mm}
\end{subfigure}
\caption{CDFs of the localization error and localization time for different techniques and scenarios. }
\label{fig:cdf}
\end{figure*}

\subsection{Analysis of RSS for Different Scenarios}
Using the simulation assumptions described earlier, the RSS observations at different UAV altitudes, which are to be used for different SSSL algorithms, are summarized in Fig.~\ref{fig:simulMeasure}. The theoretical RSS is obtained based on~\eqref{eq:7}, while the measured RSS is also derived from the same equation but includes a Gaussian noise component. Figs.~\ref{fig:rssOmni}-\ref{fig:rssDipole70} depict the variation of RSS as a function of the distance between the transmitter and the receiver.
Figs.~\ref{fig:rssTrajOmni}-\ref{fig:rssTrajDipole70} represent RSS in relation to the UAV's trajectory. In these figures, RSS at close distances to the target location is significantly influenced by the UAV's altitude. Using these measurements, the proposed localization algorithms are executed and compared across different UAV altitudes in the following section. 

\subsection{Analysis of Distance Estimation Accuracy}

In this section, we analyze how the accuracy of distance and location estimates change with respect to the true distance and angle between the UAV and the ground target. The root mean square error (RMSE) of the distance estimates using~\eqref{eq:9} as a function of the distance (DROD) and angle (DROA) between the UAV and the ground target are depicted in Figs.~\ref{fig:distrmseOmni}-\ref{fig:angDistRmseDipole70}. In addition, the RMSE of the location estimates using~\eqref{eq:13} as a function of the distance (LROD) and angle (LROA) between the UAV and the ground target are illustrated in Figs.~\ref{fig:distLocalRmseOmni}-\ref{fig:angLocalRmseDipole70}. The RMSE of the estimated distance has proportional relations with the real distance for both omnidirectional and dipole antenna patterns when the altitude of the UAV is below 50 meters. On the other hand, when the UAV is at an altitude of 70 meters, the close distances also result in poor estimation due to weaker signal strength caused by the dipole antenna pattern. Similar trends in the relationship between distance estimation RMSE and angle are also shown. Assuming that the altitude of the UAV is fixed in each experiment, the angle variation is mainly affected by the distance. In the figures, larger angles exhibit lower RMSE on distance estimates for both antenna patterns when the UAV altitude is below 50 meters. However, the distance estimation RMSE increases at higher angles when the UAV is at an altitude of 70 meters. 

The impact of angle and distance variations on localization performance is also illustrated in Fig.~\ref{fig:rmselMeasure}. The angle and distance differ at every trajectory index, so tracking the characteristics of each index is crucial. From this perspective, the LLS-CHLM algorithm is implemented for localization results in Fig.~\ref{fig:rmselMeasure}. Recalling the features of the other two algorithms: 1) the LLS-CLS only uses the closest five indices, and 2) the LLS-CUM uses every index up to the current flight index, they cannot explicitly represent the influence of each index. In addition, to clearly observe the variation in distance and angle, the transmitter is positioned at the center of the UAV's trajectory. In Fig.~\ref{fig:distLocalRmseOmni}-\ref{fig:distLocalRmseDipole70}, it is observed in all antenna patterns and UAV altitudes that larger distances cause higher localization RMSE. On the other hand, higher angles result in a lower localization RMSE. One remarkable observation is the discrepancy in how the distance and localization errors depend on angle when the UAV altitude is at 70 meters. In Fig.~\ref{fig:angDistRmseDipole70}, the distance RMSE increased when the angle is larger than 60 degrees. On the other hand, in Fig.~\ref{fig:angLocalRmseDipole70}, the localization RMSE is stabilized after 30 degrees. In this case, the possible localization error caused by the distance error is mitigated by other indices when using the least-square localization algorithm.

\subsection{Analysis of Proposed SSSL Methods}
Utilizing the estimated distances and least-square approach, we implemented and compared the proposed localization algorithms for the omnidirectional antenna pattern at a UAV altitude of 30 meters and the dipole antenna pattern at UAV altitudes of 30, 50, and 70 meters, as depicted in Fig.~\ref{fig:compLocal}. In this figure, the x-axis represents the distance traveled by the UAV, while the y-axis indicates the localization RMSE. For performance comparison, we set a localization RMSE threshold at 20 meters. For on-target location setup, the LLS-CUM achieves the localization RMSE threshold in the shortest flight time. However, for mid-target and far-target setups, the LLS-CHLM and LLS-CUM exhibit similar localization performance for both dipole and omnidirectional antenna configurations, while the LLS-CLS demonstrates unreliable performance, particularly in far-target settings.

Fig.~\ref{fig:cdf} presents the long-term localization RMSE for each algorithm with respect to target variation and the required flight time to reach the RMSE threshold in the CDF graphs. For simplicity and to distinguish the antenna pattern's impact, we primarily consider the dipole antenna pattern with a UAV altitude of 50 meters. The LLS-CLS delivers the best localization accuracy in on-target (1.8814 meters) and mid-target (2.6457 meters) scenarios. However, the same algorithm exhibits the worst performance for the far-target scenario, as the UAV's trajectory does not pass nearby the target. Both the LLS-CUM and LLS-CHLM demonstrate relatively similar localization accuracy in every target setup. For the required flight time measurement, we assume a maximum UAV flight speed of \SI{20}{\meter\second} and an acceleration speed of \SI{5}{\metre\per\square\second}. The required flight time for LLS-CLS is heavily influenced by the target's position, with the current target configuration showing the longest flight time for the required localization accuracy. The other two algorithms display similar performance in terms of required time, with LLS-CUM requiring slightly less flight time than LLS-CHLM in the overall target configurations. Table \ref{table1} summarizes the performances of all SSSL approaches, considering different evaluation criteria.

\section{Conclusions}\label{Sec:Conclusion}
In this study, we study the SSSL problem with an autonomous UAV, considering three distinct approaches for selecting anchor locations along a UAV's trajectory. We incorporate a two-ray propagation model and specific UAV antenna patterns into our simulation setup. We demonstrate that the LLS-CLS algorithm excels in terms of localization accuracy when the target is in close proximity to the UAV's trajectory. However, it falls short of achieving the required localization accuracy when the target is far from the trajectory. In contrast, the other two algorithms, LLS-CUM and LLS-CHLM, display similar and relatively stable localization performance in terms of both accuracy and required search time. Nonetheless, the LLS-CUM algorithm entails high computational complexity due to the expanding size of parameter matrices used for location estimation. From our analysis, the LLS-CHLM algorithm emerges as a promising choice, offering a balanced performance trade-off between accuracy and search time with comparatively low computational complexity. Looking ahead, our future work will examine these algorithms under realistic noise conditions. Moreover, we plan to test the proposed approaches in a real-world UAV testbed, utilizing the emulation and testbed environments of the NSF AERPAW platform at NC State University. 

\begin{table}[t]\label{t}
\caption{SSSL algorithms performance summary.}
\label{table1}
\begin{center}
\scalebox{0.68}{
\begin{tabular}{|p{0.7in}|p{0.4in}|p{0.5in}|p{1.5in}|p{0.5in}|}
\hline
\textbf{Algorithms} & \multicolumn{1}{c|}{\textbf{\begin{tabular}[c]{@{}c@{}}UAV flight \\ time (seconds)\end{tabular}}} & \multicolumn{1}{c|}{\textbf{\begin{tabular}[c]{@{}c@{}}UAV flight \\ distances (m)\end{tabular}}} & \multicolumn{1}{c|}{\textbf{\begin{tabular}[c]{@{}c@{}}Average long term \\ accuracy (m) (Var (m$^2$))\end{tabular}}} \\ \hline
CUM-On                 & 35.9                                                                   & 683.6                                                                            & 2.0058 (1.5653)                                                  \\ \hline
CHLM-On                & 77.3                                                                   & 1,369.1                                                                            & 5.8384 (19.5932)                                                  \\ \hline
CLS-On                 & 142.9                                                                   & 2,462.6                                                                          & 1.8814 (1.6371)                                                   \\ \hline
CUM-Mid                & 156.6                                                                  & 2,782.5                                                                            & 11.2024 (77.2772)                                                   \\ \hline
CHLM-Mid               & 125.8                                                                   & 2,192.2                                                                          & 7.7171 (20.4467)                                                   \\ \hline
CLS-Mid                & 238.6                                                                 & 4,142.6                                                                            & 2.6457 (3.0880)                                                  \\ \hline
CUM-Far                & 240.4                                                                 & 4,190.5                                                                          & 14.2262 (124.4173)                                                   \\ \hline
CHLM-Far               & 244.6                                                                   & 4,225.9                                                                          & 15.6131 (100.1289)                                                     \\ \hline
CLS-Far                & $-$                                                                   & $-$                                                                          & 85.8901 (2,883.2)                                                     \\ \hline
\end{tabular}}
\end{center}
\vspace{-5mm}
\end{table}

\bibliographystyle{IEEEtran}
\bibliography{references}

\vspace{12pt}

\end{document}